\documentclass[
aps
,groupedaddress
,amsfonts,amsmath,floatfix,eqsecnum,amssymb
,nofootinbib
,twocolumn
]{revtex4}

\usepackage{mathtools}
\usepackage{hyperref}
\usepackage{mathrsfs}
\usepackage[utf8]{inputenc}
\usepackage{epsfig,graphicx}
\usepackage{bm}
\usepackage{amsmath,amsfonts}
\usepackage{enumerate}
\usepackage{ytableau}
\usepackage{cancel}
\usepackage{multirow}
\usepackage{mathptmx}      % use Times fonts if available on your TeX system
\usepackage{latexsym}
\usepackage{amstext}
\usepackage{array}
\usepackage[T1]{fontenc} % if needed

\DeclareMathSymbol{\gtrless} {\mathrel}{AMSa}{"3F}

\newcommand{\ignore}[1]{}
\newcommand{\nec}[1]{\eqref{eq:#1}}

\newcommand{\eq}[1]{eq. \eqref{eq:#1}}
\newcommand{\eqs}[1]{eqs. \eqref{eq:#1}}
\newcommand{\Eq}[1]{Eq. \eqref{eq:#1}}

\newcommand{\be}{\begin{equation}}
\newcommand{\ee}{\end{equation}}
\def\bes#1\ees{%
  \begin{equation}
    \begin{split}
      #1
    \end{split}
  \end{equation}
}
\def\bs#1\es{%
    \begin{split}
      #1
    \end{split}
}
\newcommand{\C}{\mathbb{C}}

\newcommand{\R}{\mathbb{R}}
\newcommand{\Z}{\mathbb{Z}}

\def\slashchar#1{\setbox0=\hbox{$#1$}
   \dimen0=\wd0 \setbox1=\hbox{/} \dimen1=\wd1
   \ifdim\dimen0>\dimen1 \rlap{\hbox to \dimen0{\hfil/\hfil}} #1
   \else  \rlap{\hbox to \dimen1{\hfil$#1$\hfil}} / \fi}

\newcommand{\fin}{\hbox{~$\diamondsuit$}}

\newcommand{\ben}{\begin{enumerate}}
\newcommand{\een}{\end{enumerate}}

\newcommand{\ds}{\displaystyle}

\DeclareMathOperator{\Tr}{Tr}

\DeclareMathOperator{\Prob}{Prob}

\newcommand{\cH}{{\mathcal H }}

\newcommand{\cK}{{\mathcal K }}

\newcommand{\cR}{{\mathcal R }}
\newcommand{\cS}{{\mathcal S }}

\newcommand{\cV}{{\mathcal V }}
\newcommand{\cW}{{\mathcal W }}

\newcommand{\vphi}{{\mbm \phi}}

\newcommand{\vsigma}{{\mbm \sigma}}

\newcommand{\vpsi}{{\mbm \psi}}

\newcommand{\vrho}{{\mbm \rho}}

\newcommand{\esp}[1]{\langle #1 \rangle}

\newcommand{\miem}[1]{\textbf{\textsf{#1}}}

\renewcommand{\ker}{\,\operatorname{ker}\,}

\DeclareMathOperator{\ran}{ran}

\newcommand{\PM}[1]{ \begin{pmatrix} #1 \end{pmatrix} }
\newcommand{\Oplus}{\ensuremath{\vcenter{\hbox{\scalebox{1.1}{$\bigoplus$}}}}}

\ignore{$}
\newcommand{\mbm}{\vec}

\newcommand{\mspan}{ \mathrm{span} }
\newcommand{\rank}{ \mathrm{rank} }

\newcommand{\fB}{\mathrm{B}}
\newcommand{\fC}{\mathrm{C}}
\newcommand{\fTB}{\mathrm{TB}}
\newcommand{\fTC}{\mathrm{TC}}
\newcommand{\fP}{\mathrm{P}}
\newcommand{\fTP}{\mathrm{TP}}
\newcommand{\dplus}{\,\dot{+}\,}
\newcommand{\tplus}{\text{+}}

\newcommand{\ket}[1]{ | #1 \rangle }

\newcommand{\ketbra}[1]{ | #1 \rangle \langle  #1 |  }

\renewcommand{\Omega}{\varOmega}

\newcommand{\tP}{\tilde{P}}

\newcommand{\tcV}{\tilde{\cV}}

\newcommand{\qed}{ \hfill $\square$ }
\renewcommand{\fin}{ \hfill $\diamondsuit$ }
\newcommand{\proof}{ \miem{Proof} }

\newcounter{mycounter}[section] % El contador se reinicia con cada sección
\renewcommand{\themycounter}{\arabic{section}.\arabic{mycounter}}

\renewcommand{\newtheorem}[1]{
  \refstepcounter{mycounter}\label{#1} \miem{Theorem \themycounter } }
\newcommand{\newlemma}[1]{
  \refstepcounter{mycounter}\label{#1} \miem{Lemma \themycounter } }
\newcommand{\newproposition}[1]{
  \refstepcounter{mycounter}\label{#1} \miem{Proposition \themycounter } }

\newcommand{\newdefinition}[1]{
  \refstepcounter{mycounter}\label{#1} \miem{Definition \themycounter } }
\newcommand{\newcorollary}[1]{
  \refstepcounter{mycounter}\label{#1} \miem{Corollary \themycounter } }

\begin{document}

\title{\textsf{
    % Decomposition of quantum states along subspaces
    Localization of quantum states within subspaces
    %{\color{blue}
      from the Lebesgue decomposition %of positive operators
     % }
  }}

\author{L. L. Salcedo}
\email{salcedo@ugr.es}

\affiliation{Departamento de F\'{\i}sica At\'omica, Molecular y Nuclear and \\
  Instituto Carlos I de F\'{\i}sica Te\'orica y Computacional, \\ Universidad
  de Granada, E-18071 Granada, Spain.
}

\date{\today}
%\sf

\begin{abstract}
  \ignore{ A precise definition is proposed for the localization probability
    of a quantum state within a given subspace of the full Hilbert space of a
    quantum system. The corresponding localized component of the state is
    explicitly identified, and several mathematical properties are
    established. Applications and interpretations in the context of quantum
    information are also discussed.  }
  Relying on the Lebesgue-type decomposition of positive operators,
  this work introduces a rigorous notion of localization probability
  of a quantum state within a given subspace of its Hilbert space. A
  non‑negative operator $A$ is uniquely decomposed as $A=B+C$, where
  $B$ is the maximal positive operator supported inside the chosen
  subspace and $C$ has support disjoint from it. The localized
  component $B$ can be expressed via the Schur complement and
  characterized through an $A$‑dependent inner product and suitable
  trace inequalities. For quantum states, this yields a probability
  $\lambda$ that a state $\rho$ be completely contained in a subspace,
  which is strictly more restrictive than the usual overlap
  probability $\Tr(P\rho)$ and enjoys concavity and super‑additivity
  properties. The resulting framework admits natural interpretations
  in quantum information, including entropic aspects and a potential
  cryptographic masking scheme based on the uniqueness of the
  decomposition.

\end{abstract}

%\pacs{ 05.10.Ln, 02.70.-c, 02.70.Ss, 12.38.Gc }

%\date{\today}

\maketitle
\flushbottom
\setlength{\unitlength}{1mm}

\tableofcontents

\newpage

\setcounter{mycounter}{0} % Reiniciamos el contador al inicio de la sección
\section{ Introduction }
\label{sec:0}

In quantum information theory, subspaces of the system Hilbert space play a
central role in a variety of contexts, including error-correcting codes,
decoherence-free or noiseless subspaces, and more general protected or “dark”
subspaces for noisy dynamics \cite{Bennet,Knill,Duan:1997pe,Zanardi:1997vm%
  ,Lidar:1998hs,Shabani:2005ssn}. In these settings one often singles out a
subspace $\cV\subseteq \cH$ that carries the relevant logical or protected
degrees of freedom, and is therefore interested in quantifying to what extent
a given state or process is effectively confined to $\cV$ rather than
exploring the full space.  Closely related ideas appear in active-space
methods and low-energy subspace approximations in many-body physics, where one
focuses on a distinguished subspace and treats the remaining dimensions as an
environment or inactive sector
\cite{Bauman:2019xqo,Kanasugi:2024ivt,Kwao:2025dut}.%
%\cite{[active-space / simulation references]}.

Subspace-oriented approaches have been developed in several directions. One
line of work characterizes completely positive maps that preserve a given
subspace or act locally with respect to a fixed decomposition of the Hilbert
space into subspaces \cite{Aberg:2003fxu,Yamasaki:2021rlb}.
Another line analyzes protected or dark subspaces for noisy quantum channels,
providing conditions under which information encoded in a subspace is immune
to certain classes of errors \cite{Majgier:2009fji,Guedes:2012ymq}.
% ,\cite{[Maassen et al. / protected subspaces references]}).
These contributions focus primarily on
dynamical aspects: how channels act on subspaces and how to engineer subspaces
with desirable stability properties.
%
%{\color{blue}
In contrast, the present work will be
concerned with a purely operator-theoretic question: for a given non-negative
operator $A\ge 0$ and subspace $\cV$, can we decompose $A$ in a unique and
canonical way into a part supported in $\cV$ and a part supported outside
$\cV$?
%}

At the same time, many quantities in quantum information are naturally
formulated as convex optimization problems over decompositions of a state into
pure or mixed components subject to constraints. Standard examples include
convex-roof constructions of entanglement and coherence measures, as well as
gauge-based formulations of general quantum resource measures
\cite{Toth:2014fwg,Regula:2017mvm,Regula:2015itx}.
% \cite{[convex-roof / resource-theory references, e.g. Regula]}.
In such settings one often maximizes or minimizes linear or convex functionals
over all decompositions $\rho = \sum_k p_k \rho_k$ satisfying structural
conditions on the components $\rho_k$.
%
%{\color{blue}
  The problem to be analyzed in the present work can be seen
as a particular instance of this general paradigm: we seek the maximal weight
that can be concentrated in a convex combination of states whose support is
contained in a given subspace $\cV$.
%}

The structure of quantum states as mixtures of pure states is at the heart of
quantum theory and quantum information. A given density operator $\rho$ admits
many distinct convex decompositions into pure states, and these decompositions
are generally highly non-unique unless $\rho$ itself is a pure
state \cite{Schrodinger:2008opj,Hughston:1993wha,Jaynes:1957zz%
  ,Nielsen:2012yss,Kirkpatrick:2005yqc}. This raises natural questions about how to quantify the
contribution of particular pure states or families of states to a given mixed
state, and how such contributions depend on the geometry of the underlying
Hilbert space.

From this perspective, a basic question arises: given a state $\rho$ and a
subspace $\cV\subseteq \cH$ with some operational meaning, what is the largest
weight with which $\rho$ can be expressed as a mixture of states supported
entirely in $\cV$? Equivalently, can one define in a precise and intrinsic way
the “component” of $\rho$ that is located in $\cV$, and the corresponding
probability that the system is completely included within $\cV$ in some convex
decomposition of $\rho$? Standard overlap probabilities of the form
$p= \Tr(P\rho)$, where $P$ is the orthogonal projector onto $\cV$, answer a
different question: they measure the probability that a projective measurement
finds the system in $\cV$, not the maximal weight of a state whose support is
entirely contained in $\cV$.

In this work we address these questions through the mathematical tool
known as Lebesgue decomposition of positive operators
\cite{Anderson:1975, Ando:1976, Kosaki:1984} (also chapter 5 of
\cite{Zhang:2005}):

\medskip
\newtheorem{th:intro} [Anderson and Trapp \cite{Anderson:1975}]~{\em
  Given a non-negative bounded operator $A$ in a Hilbert space $\cH$,
  and $\cV$ a closed subspace of $\cH$, there exist unique
  non-negative operators $B$ and $C$ such that
\be
A = B + C
\ee
fulfilling the conditions $\ran(B^{1/2})\subseteq \cV$ and
$\ran(C^{1/2})\cap \cV=0$. Furthermore, $B$ is the maximum
non-negative operator such that $B\le A$ and $\ran(B)\subseteq
\cV$.
}

\medskip
$B$ is called the {\em shorted operator} (of $A$ with respect to
$\cV$) in \cite{Anderson:1975}, and in the finite dimensional case is
the Schur complement of $A$ with respect to $\cV^\perp$.  Explicit
constructions and proofs are provided below for the finite-dimensional
case.

The properties of the Lebesgue decomposition of positive operators
allow us to interpret $B$ as the component of $A$ along $\cV$, and to
define a localization weight $\lambda= \Tr(B)$ which can be viewed as
the maximal probability with which the system can be completely
localized within $\cV$ in any convex decomposition of the state. We
establish several structural properties of this construction,
including concavity of $B$ and of $\lambda$ as functions of $A$, trace
inequalities relating $\lambda$ to the standard overlap probability
$p=\Tr(P\rho)$, and behavior under tensor products and direct sums of
systems.
 
The rest of the paper is organized as follows: In Sec. II we exploit
(relying on results to be presented later) the mathematical properties
of the Lebesgue decomposition as regards to quantum states and discuss
the interpretation of $\lambda$ as an inclusion probability, its
relation to the usual overlap probability, its concavity and
pseudo-entropy-like features, and its behavior for composite
systems. We also illustrate how localization probabilities can be
combined and how they give rise to inequalities reminiscent of those
appearing in information theory. Lastly, we outline a simple
cryptographic scenario in which the localization probability plays a
central role.  In Sec. III we discuss the mathematical problem in the
general setting of non-negative operators and subspaces although
restricted to the finite-dimensional case, providing the proofs of the
result analyzed in the previous section. The existence and uniqueness
of the decomposition is proven, as is its relation to the Schur
decomposition, besides many auxiliary properties are established. Some
of these properties might admit an extension beyond the
finite-dimensional case. We summarize our conclusions in Sec. IV.

%%%%%%%%%%%%%%%%%%%%%%%%%%%%%%%%%%%%%

\section{ Quantum information }
\label{sec:2}

As is well known, a quantum state $\rho$ can be expressed in many different
ways as a mixture of pure states, unless the state $\rho$ itself is a pure
state. Given a set of pure states $\{ \psi_k\}_{k=1}^n$, the question then
arises: when is it possible to obtain $\rho$ as a mixture including those
states with a non-vanishing probability. A necessary and sufficient condition
for this to be possible is that all the $\psi_k$ must belong to the range
(image) of the operator $\rho$ (Proposition \ref{pr:6} below).

A further question is: in which mixtures do those selected pure states have
the largest weight? The question is well-posed in the case of $n=1$: which is
the maximum weight a pure state $\psi$ can have when it appears in a mixture
yielding $\rho$? That is, if
\be
\rho = \lambda \ketbra{\psi} + (1-\lambda) \rho',
\ee
what is the maximum value the probability $\lambda$ can attain? Clearly, if
the state $\rho'$ itself can be expressed as
\be
\rho' = \lambda' \ketbra{\psi} + (1-\lambda') \rho''
\ee
with $\lambda'>0$, the original weight $\lambda$ can be increased to a value
$\lambda + \lambda'$. The maximum probability will then be attained when
$\lambda'=0$, that is, when $\psi \not\in \ran(\rho')$.

\begin{figure}[ht]
  \begin{center}
    \includegraphics[height=50mm]{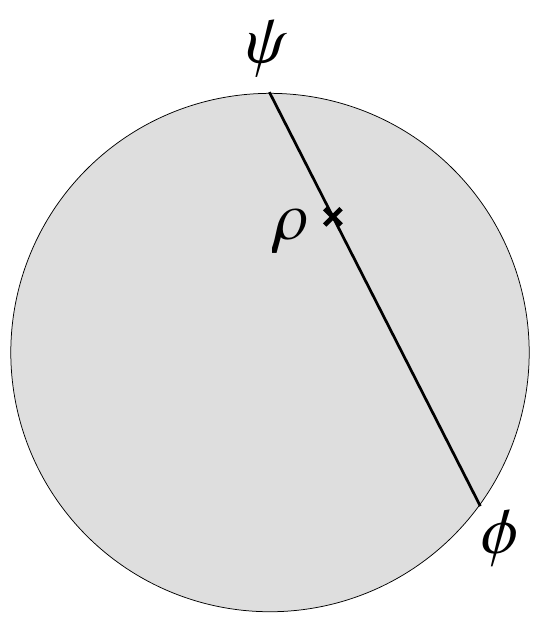}
    \end{center}
    \caption{Geometric decomposition of a qubit state $\rho$ in the Bloch
      sphere (graphically simplified to a disk).  The state is expressed as
      a mixture (convex combination or interpolation) of two pure states:
      $\rho= \lambda \ketbra{\psi} + (1-\lambda) \ketbra{\phi}$. For any given
      state $\rho$ and given pure state $\psi$, the weight $\lambda$ is
      univocally determined, as is the pure state $\phi$ (unless
      $\lambda=1$). The distances from $\rho$ to $\psi$ and $\phi$ are
      proportional to $1-\lambda$ and $\lambda$, respectively.}
\label{fig:2}
\end{figure}

The simplest instance is that of a qubit in a state $\rho$ to be expressed as
a mixture necessarily involving a certain pure state $\psi$ plus a remainder
$\rho'$. For a qubit, the condition $\psi \not\in \ran(\rho')$ implies that
$\rho'$ must be a certain pure state $\phi$, and different from
$\psi$. (Exceptionally, when $\rho$ is precisely $\psi$, $\lambda=1$ and the
state $\rho'$ is undefined.)  That is
\be
\rho = \lambda \ketbra{\psi} + (1-\lambda) \ketbra{\phi}, \quad 0 \le
\lambda \le 1
\,.
\ee
Here, $\rho$ and $\psi$ are given inputs, while $\lambda$ and $\phi$ are
outputs determined by the construction. The geometric construction in
Fig. \ref{fig:2} illustrates the solution.  Except in the trivial case
$\rho= \ketbra{\psi}$, the state $\phi$ is uniquely defined. $\rho$ is a point
of the Bloch closed ball, with associated vector $\vrho$,
~$\rho= \frac{1}{2}(I+\vrho\cdot\vsigma)$, with $\|\vrho\|\le1$. The pure
state $\psi$ corresponds to a normalized vector $\vpsi$ on the sphere. The
state $\phi$ corresponds to the interpolation
~$\vrho = \lambda \vpsi + (1-\lambda) \vphi$, ~or rather extrapolation since
$\vrho$ and $\vpsi$ are given while $\lambda$ and $\vphi$ result from the
construction.

To go beyond the single-qubit and $n=1$ cases, we analyze the following
problem: given a quantum state $\rho_A$ in a Hilbert space $\cH$ and a certain
subspace $\cV \subseteq \cH$, to what extent can the state be obtained as a
mixture (a convex combination) of a state $\rho_B$ located in $\cV$ and a
state $\rho_C$ located outside $\cV$,
~$\rho_A = \lambda \rho_B + (1-\lambda) \rho_C, \quad 0\le \lambda \le
1$. ~Here by $\rho_B$ to be ``located'' in $\cV$ we mean that it is a mixture
of pure states in $\cV$.  As is well-known (Lemma \ref{lm:0} below), a state
$\rho$ can be expressed as a mixture of pure states in many ways, but those
pure states must belong to the image of the operator $\rho$, and conversely,
the linear span of the set of pure states in the mixture must be the image of
$\rho$. Thus, the condition on $\rho_B$ (to be located in $\cV$) can be
expressed as ~$\ran(\rho_B) \subseteq \cV$, while the condition on $\rho_C$,
expressing that it is located outside $\cV$, will be formulated as
~$\ran(\rho_C) \cap \cV = 0$. Having delimited the mathematical problem, we
will study the decomposition
\bes
\rho_A &= \lambda \rho_B + (1-\lambda) \rho_C, \qquad 0 \le \lambda \le 1
\,,
\\
\quad
& \ran(\rho_B) \subseteq \cV
,
\quad
\ran(\rho_C) \cap \cV = 0
.
\label{eq:1.2}
\ees
Here, the data are $\rho_A$ and $\cV$; the weight $\lambda$ and the states
$\rho_B$ and $\rho_C$ are not chosen a priori, but are a determined by the
decomposition itself.

\subsection{ Support of a quantum state }
\label{sec:2.1}

The meaning and significance of the support (image) $\ran(\rho)$ of a quantum
state $\rho$ ~($\rho\ge 0$ ~and ~$\Tr(\rho)=1$) ~follow from the following
lemma and proposition:

\medskip \medskip%
\newlemma{lm:0} ~{\em Given a set of vectors ~$\{ \tilde\psi_k\}_{k=1}^n$
\be
\ran\left( \sum_{k=1}^n  \ketbra{\tilde\psi_k} \right) = \mspan\{ \tilde\psi_k\}_{k=1}^n
\,.
\ee
}

\medskip %
\proof ~Given another set vectors, $\{ \tilde\phi_k\}_{k=1}^n$, the following
equality
\be
\sum_{k=1}^n  \ketbra{\tilde\psi_k}
=
\sum_{k=1}^n  \ketbra{\tilde\phi_k} =:A
\ee
holds if and only if the two sets are related through a unitary
transformation. The proof of this can be found in \cite{Nielsen:2012yss},
  p.~103. Clearly, ~$\ds \ran(A) \subseteq \mspan\{
\tilde\psi_k\}_{k=1}^n$. ~On the other hand, taking the $\tilde\phi_k$ as the
eigenvectors of $A$, each $\tilde\psi_k$ is a (unitary) linear combination of
such eigenvectors; hence, $\tilde\psi_k \in \ran(A) $.  \qed

\medskip \medskip %
\newproposition{pr:6} ~{\em Let $\rho$ be a quantum state, and let
  ~$\{ \psi_k\}_{k=1}^n$ ~be a finite set of unit vectors (pure states). The
  condition ~$\forall k ~\psi_k\in\ran(\rho)$ ~is necessary and sufficient for
  the existence of a mixture yielding $\rho$ where each of the $\psi_k$ is
  present with a non-vanishing weight.  }

\medskip %
\proof ~The condition $\psi_k\in \ran(\rho)$ is necessary by Lemma \ref{lm:0}.

To prove its sufficiency, let us express the state as a mixture composed of
its normalized eigenvectors,
\be
\rho = \sum_j \lambda_j \ketbra{\varphi_j}
,
\quad
\lambda_j > 0, \quad\sum_j \lambda_j =1
.
\ee
Let $\lambda_m$ denote the smallest non-zero eigenvalue, then
~$\rho \ge \lambda_m P$, ~where $P=\sum_j\ketbra{\varphi_j}$ is the orthogonal
projector onto $\ran(\rho)$.  Because $P$ is the identity operator within the
subspace $\ran(\rho)$, it can be expressed as a sum of projectors associated
to an orthonormal basis of $\ran(\rho)$, and for each $k$, such a basis can be
chosen so that $\psi_k$ is one of its elements, $P=\ketbra{\psi_k}+
\cdots$. Therefore
\be
\rho \ge \frac{\lambda_m}{n} n P =
\frac{\lambda_m}{n} \sum_{k=1}^n ( \ketbra{\psi_k}+ \cdots )
.
\ee
Thus each $\psi_k$ appears in the mixture of $\rho$ with weight at least
$\lambda_m/n >0$.  \qed

\subsection{ Decomposition of a quantum state along a subspace }
\label{sec:2.2}

\medskip \medskip%
The mathematical results to be developed in the Sec. \ref{sec:1}
immediately apply to the decomposition of a quantum state $\rho_A$
along a subspace $\cV\subseteq\cH$, as defined in \eq{1.2}.  Using the
definitions introduced in \eq{2.16}
\bes
\lambda \rho_B &= \fB(\rho_A | \cV) ,\qquad
\lambda = \fTB(\rho_A | \cV), \\
(1-\lambda) \rho_C &= \fC(\rho_A | \cV) .
\ees

If, for simplicity, $\rho_A >0$ is assumed and the density matrix is expressed
in block form with respect to $\cH=\cV \oplus\cV^\perp$ as
\be
\rho_A = \PM{ a & b^\dagger  \\ b & c }
\label{eq:3.5}
,
\ee
then
\be
\lambda \rho_B = a - b^\dagger c^{-1} b
,
\label{eq:3.5a}
\ee
that is, $\lambda \rho_B$ is the Schur complement of the block $c$ in the
matrix $\rho_A$.

The existence and uniqueness of the decomposition in \eq{1.2} is the content
of the Theorem \ref{th:1} in Sec. \ref{sec:1a}. In the particular cases
$\lambda=1$ or $\lambda=0$, the states $\rho_C$ or $\rho_B$ are undetermined,
respectively, but $\lambda$ itself is always well-defined.

In view of the meaning of the support of a state, the condition on $\rho_B$
has a clear interpretation: ~$\ran(\rho_B) \subseteq \cV$ ~ensures that all
the pure states in a mixture producing $\rho_B$ belong to $\cV$.

The condition ~$\ran(\rho_C) \cap \cV = 0$ ~is intended to ensure that $\lambda$
is maximal in a decomposition constrained by $\ran(\rho_B) \subseteq
\cV$. That $\lambda$ is indeed maximal when $\ran(\rho_C) \cap \cV = 0$ is an
immediate consequence of the Lemma \ref{lm:1} or \eqs{2.2}. Therefore, the
decomposition in \eq{1.2} of a state $\rho_A$ yields the maximal probability
for the states $\rho_B$ lying in a given subspace $\cV$.

According to the Proposition \ref{pr:10}, clause 4, the decomposition enjoys a
reciprocity property, namely, if the subspace $\cW:=\ran(\rho_c)$ is chosen a
priori, \eq{1.2} is also the decomposition of $\rho_A$ requiring the support
of $\rho_C$ to be in $\cW$ and $\ran(\rho_B)\cap \cW=0$.

For given $\rho_A\ge 0$ and $\cV$, the decomposition takes place on the
support of $\rho_A$. The subspace ~$\tcV:=\ran(\rho_A) \cap \cV$ ~is the
support of $\rho_B$ ~(Theorem \ref{th:4}), and $\lambda>0$ unless $\tcV=0$.
Hence, in particular, $\lambda>0$ when $\rho_A>0$.\footnote{Assuming
  $\cV\neq 0$.} ~The support of $\rho_C$ is also contained in $\ran(\rho_A)$,
and the (not necessarily orthogonal) direct sum of the supports of
$\rho_B$ and $\rho_C$ yields that of $\rho_A$. The decomposition of
$\rho_A\ge 0$ along $\cV$  is identical to that of
$\rho_A>0$ on $\ran(\rho_A)$ along $\tcV$ ~(Corollary \ref{co:4}).

It should be stressed that in general, the subspace $\ran(\rho_C)$ is not
orthogonal to $\cV$. This is also clear in the example of the qubit in
Fig. \ref{fig:2} since in the Bloch sphere, two orthogonal pure states
correspond to diametrically opposed points. Orthogonality is ensured when
$\tcV$ is an invariant subspace of the state $\rho_A$ (clause 5 of the
Proposition \ref{pr:10}), and in this case, the states $\rho_B$ and $\rho_C$
have zero (orthogonal) overlap.

When $\rho_A>0$ a new scalar product can be defined in $\cH$, to wit,
\be
(\psi|\phi):=\esp{\psi|\rho_A^{-1}|\phi}
\label{eq:3.7}
, 
\ee
such that the support of $\rho_C$ is just the orthogonal complement of $\cV$.
Moreover the operators ~$\lambda \rho_B \rho_A^{-1}$ ~and
~$(1-\lambda) \rho_C \rho_A^{-1}$ ~are precisely the projectors onto the two
subspaces (\eq{2.25}). Of course, this implies the relation
\be
\rho_B \rho_A^{-1} \rho_C = 0
\ee
provided $\cV$ differs from $0$ and $\cH$, so that $\lambda(1-\lambda)\neq0$ 
and $\rho_B$ and $\rho_C$ are well-defined.

Returning to the general case $\rho_A\ge 0$, also intriguing is the
following relation derived from the Proposition \ref{pr:7}:
\be
\lambda \rho_B =  (\tP \rho_A^{-1} \tP)^{-1} \,,
\ee
where $\tP$ is the orthogonal projector onto $\tcV$, the inverse of $\rho_A$
is taken within its support, and the inverse of ~$\tP \rho_A^{-1} \tP$ ~is
taken within $\tcV$. In the particular case of $\cV$ being a one-dimensional
subspace spanned by a unit vector $\psi \in \ran(\rho_A)$, the formula implies
\be
\lambda = \frac{1}{\esp{\psi|\rho_A^{-1} |\psi}}
,
\qquad
\rho_B = \ketbra{\psi}
\,.
\ee
This result is consistent with that quoted in Ref. \cite{Nielsen:2012yss},
  p.~105.

One interesting conclusion from the previous formula is that the values of
$\lambda$ obtained as $\psi$ (and so $\cV$) changes, fully characterize a
state $\rho_A$ (Proposition \ref{pr:5}).

Since a density matrix is described by $d^2-1$ real parameters (being
$d$ the dimension of the quantum Hilbert space $\cH$), it could be
expected that the $\lambda_k$ values of a suitably chosen fixed finite
set $\{\psi_k\}$ of pure states (each spanning a one-dimensional
subspace $\cV_k$) would be sufficient to univocally identify the
density matrix. In the case of a qubit, such pure states could be the
states $\ket{\tplus x}$, $\ket{\tplus y}$, and $\ket{\tplus
  z}$. Actually, this expectation is correct when $\rho_A>0$, that is,
when its support is the entire space $\cH$, but it fails when
$\rho_A \ge 0$ is allowed. For instance, if $\rho_A$ is a pure state
$\phi$, the values $\lambda_k$ will vanish for each $\psi_k$ in the
set, unless $\phi$ is one of them. The caveat is that the values
$\esp{\psi_k|\rho_A^{-1} |\psi_k}$ do characterize $\rho_A^{-1}$ but
they do not coincide with $\lambda_k^{-1}$ unless
$\psi_k\in\ran(\rho_A)$.

Further mathematical properties about the decomposition of
non-negative operators uncovered in Sec. \ref{sec:1} can be translated
to quantum states. Some of them will be discussed subsequently.

\subsection{ Interpretation of the decomposition as a probability  }
\label{sec:2.3}

Given a state $\rho_A$ and a subspace $\cV$, the standard quantum-mechanical
assignation is the (orthogonal) overlap, that is,\footnote{The various functions, $\fB$, $\fC$, etc are defined in \nec{2.16}.}
\be
p:= \fTP(\rho_A |\cV) = \Tr(P\rho_A)
,
\ee
where $P$ denotes the orthogonal projector onto $\cV$. This is the probability
of leaving the system in a state contained in $\cV$ after an ideal measurement
of the PVM ~$I=P+P^\perp$. Such probability vanishes if and only if the
support of the state $\rho_A$ is orthogonal to $\cV$.

The definition in \eq{1.2} of the quantity ~$\lambda = \fTB(\rho_A |\cV)$
~suggests a quantum-information interpretation which is different from the
overlap, namely, $\lambda$ can be regarded as the probability (or weight) of
the state $\rho_A$ to be {\em completely included} (or localized) within the
subspace $\cV$ of pure states. The inclusion probability $\lambda$ vanishes
iff the support of $\rho_A$ is disjoint (in vector-space sense) from $\cV$,
and attains its maximum value $\lambda=1$ iff the support of the state is
contained in $\cV$.

A property consistent with this interpretation is that $\lambda$ is an
increasing function of $\cV$, that is (Proposition \ref{pr:3}, clause 1)
\be
\cV_1 \subseteq \cV_2 ~\implies ~\fTB(\rho_A |\cV_1) \le \fTB(\rho_A |\cV_2)
,
\ee
and in fact, the statement is stronger: not only $\lambda_1 \le \lambda_2$ but
also
\be
\lambda_1 \rho_{B_1} \le \lambda_2 \rho_{B_2}
\ee
as operators.

The concept of {\em inclusion probability} within a subspace of pure states
is more restrictive than that of overlap, and indeed the following inequality
always holds:
\be
\lambda \le p
\label{eq:3.13}
\ee
(Corollary \ref{co:1}, clause 1).  For instance, for a qubit in state
$\ket{\tplus z}$ and subspace $\cV$ spanned by $\ket{\tplus x}$, the overlap
is 50\%, implying that the two states are partially equal and partially
different. However, the weight $\lambda$ vanishes, as $\ket{\tplus z}$ can
never be included within the subspace $\mspan(\ket{\tplus x})$, that is, if
$\ket{\tplus z}$ is written as a mixture including $\ket{\tplus x}$ the latter
must appear with zero weight.

If $\lambda$ is interpreted as the probability of inclusion of $\rho_A$ within
the subspace $\cV$, the weight $1-\lambda$ carried by $\rho_C$ must be
interpreted as the probability of the state $\rho_A$ not to be completely
included within $\cV$. As already noted, in the decomposition
\be
\rho_A = \lambda \rho_B + (1-\lambda) \rho_C
\,,
\ee
the support of $\rho_C$ is an output, not a constraint, and such support is not
necessarily orthogonal to that of $\rho_B$. This means that
quantum-mechanically, those two states $\rho_B$ and $\rho_C$ are not completely
distinguishable. They are when $\cV$ is an invariant subspace of $\rho_A$, that
is, an orthogonal direct sum of eigenspaces of $\rho_A$. In that case, the
decomposition based on inclusion coincides with that based on overlap:
\be
 [P,\rho_A]=0 \implies \rho_A = P  \rho_A P + P^\perp \rho_A P^\perp
\,,
\ee
and this is a necessary and sufficient condition for $\lambda=p$ ~(Corollary
\ref{co:1}, clause 1).

As said, the condition $\ran(\rho_B) \subseteq \cV$ is stronger than that of
overlap of $\rho_A$ with $\cV$. Correspondingly, the condition
$\ran(\rho_C) \cap \cV = 0$, which intuitively means that $\rho_C$ lies
outside $\cV$, is weaker than the condition ~$\ran(\rho_C) \perp \cV$, which
expresses the same idea in a much stronger sense. The latter condition is too
strong to be compatible with the inclusion requirement
$\ran(\rho_B) \subseteq \cV$.

Some insight is obtained by considering the complementary decomposition, along
$\cV^\perp$, that is
\bes
\rho_A &= \lambda^\perp \rho_{B^\perp} + (1-\lambda^\perp) \rho_{C^\perp},
\qquad 0 \le \lambda^\perp \le 1
\,,
\\
\quad
& \ran(\rho_{B^\perp}) \subseteq \cV^\perp
,
\quad
\ran(\rho_{C^\perp}) \cap \cV^\perp = 0
.
\label{eq:1.2b}
\ees
Here $\lambda^\perp$ is the probability of complete inclusion of the state
$\rho_A$ within the subspace $\cV^\perp$. Once again ~$\lambda^\perp \le 1-p$,
and combined with \eq{3.13},
\be
\lambda \le p \le 1-\lambda^\perp \,.
\label{eq:3.19}
\ee
According to these inequalities, the condition of complete inclusion within
$\cV$ is more restrictive than the condition of overlap, and in turn, the
overlap is more restrictive than the condition of non-complete inclusion of
$\rho_A$ within $\cV^\perp$.

\begin{figure}[t]
  \begin{center}
    \includegraphics[height=60mm]{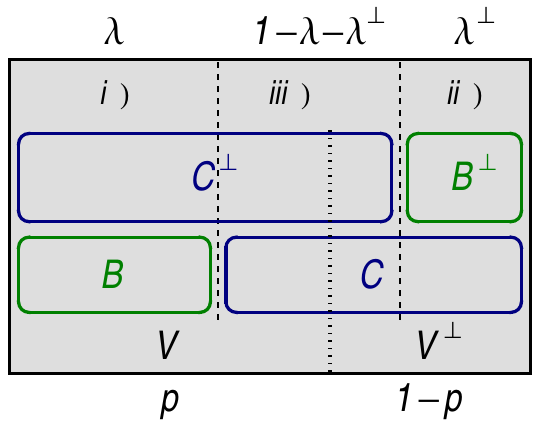}
    \end{center}
    \caption{Schematic representation of the decompositions
      $\rho_A = \lambda \rho_B + (1-\lambda) \rho_C$ along $\cV$, and
      $\rho_A = \lambda^\perp \rho_{B^\perp} + (1-\lambda^\perp)
      \rho_{C^\perp}$ along $\cV^\perp$.  The states of type $B$ and $B^\perp$
      are fully inside $\cV$ and $\cV^\perp$ respectively. The sectors $i)$ and
      $ii)$ have sizes $\lambda$ and $\lambda^\perp$, respectively.}
\label{fig:1}
\end{figure}

Quantum-mechanically, the events of overlap with $\cV$ and $\cV^\perp$ are
incompatible and the corresponding probabilities $p$ and $1-p$ add up to one.
The events of being fully within $\cV$, with weight $\lambda$, and being fully
within $\cV^\perp$, with weight $\lambda^\perp$, are also incompatible. This
is consistent with the relation
\be
\lambda + \lambda^\perp \le 1
\label{eq:3.18}
\ee
(which follows from \eq{3.19}), since the sum of two probabilities does not
exceed one. However, there is a non-negative deficiency
$1-\lambda - \lambda^\perp$ which in general is not zero. The partition
\be
1 = \lambda + \lambda^\perp  + (1-\lambda - \lambda^\perp)
\ee
suggests that for a given state $\rho_A$ there are three distinct
sectors, induced by the decomposition $\cH = \cV \oplus \cV^\perp$:
\begin{itemize}
  
\item[$i)$] States located fully within $\cV$, with size $\lambda$.

\item[$ii)$] States located fully within $\cV^\perp$, with size $\lambda^\perp$.

\item[$iii)$] States not fully located within $\cV$ nor fully within
  $\cV^\perp$, so partially on both subspaces. This sector has size
  $1-\lambda - \lambda^\perp$.

\end{itemize}
$\rho_B$ would fill up the first sector and $\rho_{B^\perp}$ the second
sector. In contrast, $\rho_C$ would fill up the union of the second and
third sectors, and $\rho_{C^\perp}$ the union of the first and third
sectors. This is illustrated in Fig.~\ref{fig:1}.

It should be stressed that, even if suggestive, the classification into the
three sectors does not extend to a corresponding decomposition of the state
$\rho_A$ itself: If the state is written as
\be
\rho_A = \lambda \rho_B + \lambda^\perp \rho_{B^\perp} + (1-\lambda
-\lambda^\perp) \rho_D,
\label{eq:3.22a}
\ee
the weight $1-\lambda -\lambda^\perp$ is non-negative and $\Tr(\rho_D)=1$, but
in general it is not true that ~$\rho_D\ge 0$, so it does not represent a
proper quantum state.  In fact, according to the Proposition \ref{pr:13},
$\rho_D$ cannot be non-negative whenever $\rho_A>0$.\footnote{If $\cV$ is an
  invariant subspace of $\rho_A$, ~$\lambda+\lambda^\perp=1$ and $\rho_D$ is
  undefined.}  Therefore, although ~$\rho_A \ge \lambda \rho_B$ ~and
~$\rho_A \ge \lambda^\perp \rho_{B^\perp}$, ~the inequality
~$\rho_A \ge \lambda \rho_B + \lambda^\perp \rho_{B^\perp}$ ~does not hold in
general.

\medskip%

\begin{figure}[t]
  \begin{center}
    \includegraphics[height=50mm]{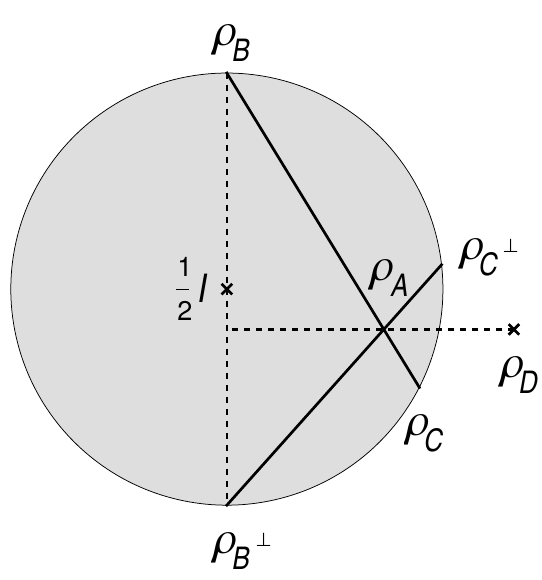}
    \end{center}
    \caption{Illustration of the decompositions of a qubit state $\rho_A$
      along one-dimensional subspaces $\cV$ and $\cV^\perp$ in the Bloch
      sphere (\eqs{3.22}--\nec{3.24}). Since $\rho_A>0$, the operator $\rho_D$
      is outside the sphere and does not define a true quantum state. That
      the straight-line joining $\rho_A$ and $\rho_D$ is a horizontal one
      illustrates \eq{3.30}, which is only granted for qubit states.}
\label{fig:3}
\end{figure}
A qubit provides an easy illustration (see Fig. \ref{fig:3}). Without loss of
generality let
\bes
\rho_A &= \frac{1}{2}(I + \vrho_A\cdot \vsigma),
\quad
\vrho_A = ( a \sin(\theta), 0, a \cos(\theta) ), 
\\
&a \in [0,1),
\qquad
\theta \in (0,\pi)
,
\label{eq:3.22}
\ees
and let $\cV$ be the subspace spanned by $\ket{\tplus z}$. In this case, one
finds that
\bes
&p = a \cos(\theta), \\
&\lambda = \frac{1}{2}\frac{1-a^2}{1-a\cos(\theta)}
,\qquad
\lambda^\perp = \frac{1}{2}\frac{1-a^2}{1+a\cos(\theta)}
,%\quad
\\
& 1-\lambda- \lambda^\perp = \frac{a^2\sin^2(\theta)}{1-a^2\cos^2(\theta)}
,
\ees
with
\bes
\vrho_B &= -\vrho_{B^\perp} = \left(0,0,1 \right),
\\
\vrho_D &= \left(
  \frac{1-a^2\cos^2(\theta)}{a\sin(\theta)},0,a\cos(\theta)
\right)
.
\label{eq:3.23}
\ees
Therefore,
\be
\|\vrho_D\|^2 = \frac{1 + (1-a^2)^2 \cot^2(\theta)}{a^2} > 1
\,.
\label{eq:3.24}
\ee
The norm is larger that one so $\rho_D$ does not represent a quantum
state; the operator $\rho_D$ has one negative eigenvalue. The case
$a=1$, which corresponds to a pure state, was excluded from the
analysis. When $a=1$, the norm is one and $\rho_D = \rho_A$, excluding
the exceptional cases $\theta\in\{0,\pi\}$, for which the pure state
is in $\cV$ or $\cV^\perp$.

\medskip \medskip%
Two different types of partitions are being considered in the previous
discussion, to wit, the partition ~$1=p+(1-p)$ ~based on $\cV$ versus
$\cV^\perp$, ~and another ~$1= \lambda + (1-\lambda)$ based on
$\rho_B$ versus $\rho_C$. The table of combined (or joint) relations
is also readily established. First, with obvious notation,
\be
\begin{aligned}
&\Prob(\cV|A) = \Tr(P \rho_A) = p, &
&\Prob(\cV^\perp|A) = 1-p
,
\\
&\Prob(B|A) = \lambda, &
&\Prob(C|A)= 1-\lambda
.
\end{aligned}
\ee
Then, using ~$\Prob(x,y|z)=\Prob(x|y,z)\Prob(y|z)$,
\bes
&\Prob(\cV|B,A)=\Tr(P \rho_B) = 1
\implies \Prob(\cV,B|A)= \lambda
\\
&\Prob(\cV^\perp|B,A) =\Tr(P^\perp \rho_B)= 0
\implies \Prob(\cV^\perp,B|A)= 0
.
\ees
Finally,
\bes
\Prob(\cV|C,A) &= \Tr(P \rho_C) =
\frac{\Tr(P (\rho_A - \lambda \rho_B) )}{1-\lambda}
\\
&= \frac{p-\lambda}{1-\lambda}
\implies \Prob(\cV,C|A) = p-\lambda
\\
\Prob(\cV^\perp|C,A) &= 1- \Prob(\cV|C,A) = \frac{1-p}{1-\lambda}
 \\ & \implies \Prob(\cV^\perp,C|A) = 1-p
  .
\ees

\begin{table}[t]
\begin{center}
\begin{tabular}{c | c  c | c }
  $A$  & $\cV$ & $\cV^\perp$ & \\
  \hline
$B$ & $\lambda$ & $0$ & $\lambda$ \\ 
$C$ & $p-\lambda$ & $1-p$ & $1-\lambda$ \\
  \hline
    & $p$ & $1-p$ & 1
\end{tabular}
\caption{Probabilities of the partition $(\cV,B)$, $(\cV^\perp,B)$,
  $(\cV,C)$, and $(\cV^\perp,C)$.}
\label{tab:1}
\end{center}
\end{table} 

The various probabilities are summarized in Table \ref{tab:1}.

\medskip%
Incidentally, comparing \eqs{3.22} and \nec{3.23}, it can be observed that
$(\vrho_D)_z = (\vrho_A)_z=p$. From \eq{3.22a}, this is equivalent to
~$p=\lambda+(1-\lambda-\lambda^\perp)p$, ~that is,
\be
p=  \frac{\lambda}{\lambda + \lambda^\perp}
\label{eq:3.30}
.
\ee
The same relation can also be expressed as
~$\Prob(C|\cV)= 1-\lambda - \lambda^\perp$, ~or also as
\be
\Prob(B|\cV) = \Prob(B^\perp|\cV^\perp)
.
\ee
If fact, such relations are valid for a qubit, or more precisely for
$\rank(\rho_A)=2$, but they do not hold in general. Indeed, using \eqs{3.5}
and \nec{3.5a},
\bes
\Prob(B|\cV) &= \frac{\lambda}{p}
= \frac{ \Tr( a-b^\dagger c^{-1} b ) }{\Tr(a)}
,
\\
\Prob(B^\perp|\cV^\perp) &= \frac{\lambda^\perp}{1-p}
= \frac{ \Tr( c-b a^{-1} b^\dagger ) }{\Tr(c)}
,
\ees
The two expressions coincide when $a,b,c$ are c-numbers, but in general, they
yield different values.

\medskip \medskip%
The probability interpretation made above can be extended to the case where a
state is decomposed into $n$ components (instead of just two). Assuming
$\rho_A >0$ for simplicity, let
\be
\rho_A = \sum_{k=1}^n \lambda_k \,\rho_{C_k}
, \qquad
\sum_{k=1}^n \lambda_k = 1
,
\label{eq:3.34}
\ee
be the decomposition of $\rho_A$ along disjoint supports:
\be
\cW_k = \ran(\rho_{C_k})
\,,
\qquad
\cH = %\dplus_{k=1}^n \cW_k
\cW_1 \dplus \cdots \dplus \cW_n
,
\label{eq:3.34a}
\ee
which are orthogonal with respect to the non-standard scalar product
introduced in \eq{3.7}. Let $P_k$ be the (standard) orthogonal projector onto
$\cW_k$, and
\be
p_k = \Tr(P_k \rho_A)
\,.
\ee
The following identifications can be made:
\be
\Prob(C_k|A) = \lambda_k,
\qquad
\Prob(\cW_k|A) = p_k,
\ee
and
\bes
\Prob(\cW_k|C_k,A) &= \Tr(P_k \rho_{C_k}) = 1,
%\qquad
\\
\Prob(\cW_k^\perp|C_k,A) &= \Tr(P_k^\perp \rho_{C_k}) =0
.
\ees
Further, the relations
\be
p_j = \sum_{k=1}^n \lambda_k  \Tr(P_j \rho_{C_k})
\ee
correspond to
\bes
\Prob(\cW_j|A) &= \sum_{k=1}^n \Prob(\cW_j|C_k,A) \Prob(C_k|A)
\\ &=
\sum_{k=1}^n \Prob(\cW_j,C_k|A)
.
\ees

\medskip \medskip%
Note that the set $\{C_k\}_{k=1}^n$ ~can be regarded as a partition, as their
probabilities add up to one, ~$\sum_{k=1}^n\lambda_k = 1$. In contrast, this
is not true for the $\cW_k$. In fact, the relation $\lambda_k \le p_k$ implies
\be
1 \le \sum_{k=1} \Prob(\cW_k|A)
,
\ee
and in general, the sum exceeds one. The $\cW_k$ may be non-orthogonal among
them, and thus quantum-mechanically, they are compatible events.

Of course, from the inequality
\be
\forall \rho_A>0 \quad \sum_{k=1} \Tr( P_k \rho_A ) \ge 1
,
\ee
it does not follow the operator identity ~$\sum_{k=1} P_k \ge I$, ~because the
operators $P_k$ depend on (are constrained by) $\rho_A$.

At variance with this, for any given $j$, $\cW_j$ and $\cW_j^\perp$ are
incompatible and the probabilities ~$\Prob(\cW_j,C_k|A)$ ~plus
~$\Prob(\cW_j^\perp,C_k|A)$ ~can be organized as a $n\times 2$ table, with
marginal probabilities ~$\lambda_k=\Prob(C_k|A)$, ~plus ~$p_j=\Prob(\cW_j|A)$
~and ~$1-p_j=\Prob(\cW_j^\perp|A)$. ~The $2\times 2$ Table \ref{tab:1}
corresponds to $n=2$, with $\cW_1=\cV$, plus $C_1=B$ and $C_2=C$. For $n>2$
the data $p_j$ and $\lambda_k$ are not sufficient to fully fix all the entries
of such $n\times 2$ table of probabilities.

\medskip \medskip%
The inequalities of the type \eq{3.18} can be much more general (and no longer
be a consequence of $\lambda\le p$). In fact, if
\be
\cV = \Oplus_{k=1}^n \cV_k
\ee
and
\bes
\rho_A &= \lambda \rho_B + (1-\lambda) \rho_C
\\ &=
\lambda_k \rho_{B_k} + (1-\lambda_k) \rho_{C_k}
\ees
are the decomposition along $\cV$ and $\cV_k$, respectively,
it holds (Theorem \ref{th:3}, clause 3),
\be
\lambda \ge \sum_{k=1}^n \lambda_k
\,.
\ee
As already said, the stronger inequality
\be
\lambda \rho_B \ge \sum_{k=1}^n \lambda_k \,\rho_{B_k}
\ee
is not guaranteed in general.

\subsection{ Entropic properties }
\label{sec:2.4}

The weighted density matrix  $\lambda\rho_B := \fB(\rho_A |\cV)$
enjoys the property (Lemma \ref{lm:1})
\be
\lambda \rho_B
 = \max\{ X ~|~ X \le \rho_A, ~~\ran(X) \subseteq \cV \}
,
\ee
and also
\be
\lambda 
 = \max\{ \Tr(X) ~|~ X \le \rho_A, ~~\ran(X) \subseteq \cV \}
.
\ee
So $\lambda = \fTB(\rho_A|\cV)$ is the maximum weight that can be concentrated
within the subspace $\cV$ in a mixture yielding $\rho_A$.

\begin{figure}[t]
  \begin{center}
    \includegraphics[height=50mm]{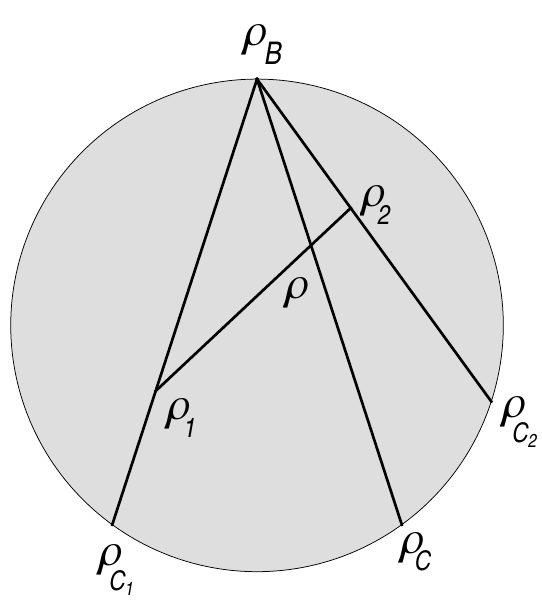}
    \end{center}
    \caption{Illustration of the concavity property for a qubit in the Bloch
      sphere. The states $\rho_1$ and $\rho_2$, and their mixture $\rho$, are
      decomposed along a one-dimensional subspace $\cV$ with orthogonal
      projector $\rho_B$. The weight $\lambda$ of $\rho$ is larger than the
      average of the weights $\lambda_1$ and $\lambda_2$ of $\rho_1$ and
      $\rho_2$.}
\label{fig:4}
\end{figure}

A consequence of the maximal property is that, for a given subspace $\cV$,
$\lambda$ is a concave function of $\rho_A$ (Theorem \ref{th:2}). That is,
if the states $\rho_{A_k}$, $k\in\{1,\ldots,n\}$, have weights $\lambda_k$ along
$\cV$, and
\be
\rho_A = \sum_{k=1}^n q_k \rho_{A_k}
\qquad q_k\ge 0, \quad \sum_{k=1}^n q_k = 1,
\ee
then
\be
\lambda \ge \sum_{k=1}^n q_k \lambda_k
,
\ee
where $\lambda$ is the weight of $\rho_A$ along $\cV$. In fact, the
statement is stronger: if $\rho_{B_k}$ are the density matrices maximally
localized on $\cV$ for each $\rho_{A_k}$, then also
\be
\lambda \rho_B \ge \sum_{k=1}^n q_k \lambda_k \rho_{B_k}
.
\ee
The concavity property is illustrated in Fig. \ref{fig:4} for a qubit.

The quantum-mechanical von Neumann entropy
$S(\rho_A) = -\Tr(\rho_A\log\rho_A)$ is a concave function, implying
that mixing tends to increase the entropy. Likewise, the concavity of
the localization weight $\fTB(\rho_A|\cV)$ indicates that it tends to
increase under mixing. Nevertheless, differences arise between the two
quantities when considering states $\rho_A$ fully localized within a
given subspace $\cV$: any such state maximizes the localization with
$\lambda=1$, while only the state $\rho_A =P/\Tr(P)$ attains a maximum
for the entropy.

Another characteristic property of the entropy is that it is additive with
respect to independent subsystems, that is, if $\rho_{A_1}$ and $\rho_{A_2}$
are two states in $\cH_1$ and $\cH_2$, respectively,
\be
S(\rho_{A_1} \otimes \rho_{A_2} )  = S(\rho_{A_1}) + S(\rho_{A_2} )
.
\ee
Such equality does not hold for the localization probability $\lambda$, rather
(Proposition \ref{pr:12})
\be
\lambda \ge \lambda_1 \lambda_2
,
\ee
where $\lambda_1$, $\lambda_2$ and $\lambda$ are the weights of the states
$\rho_{A_1}$, $\rho_{A_2}$ and $\rho_A = \rho_{A_1} \otimes \rho_{A_2}$ along
the subspaces $\cV_1$, $\cV_2$ and $\cV_1\otimes\cV_2$, respectively.

A relation which is closer to additivity is
\be
\cS \ge \cS_1 + \cS_2
\ee
with $\cS := \log(\lambda)$, although still this quantity is
super-additive rather than additive. Since the map
$\rho_A \to \lambda$ is concave and the function $\log$ is both
increasing and concave, it follows that $\rho_A \to \cS$ is also
concave, as it would be required for an entropy.

\subsection{Measurement of localization }

According to quantum mechanics, the probability $p=\Tr(P \rho_A)$ (where $P$
denotes the orthogonal projector onto a subspace $\cV$) can be determined by
doing ideal measurements of the PVM $I=P+P^\perp$ on the state $\rho_A$. In
principle, such an approach is not available to the determination of the
inclusion probability $\lambda= \fTB(A|\cV)$, in the decomposition
$\rho_A = \lambda \rho_B + (1-\lambda )\rho_C $. The reason is that the
quantity $\fTB(A|\cV)$ scales as $\fTB(\mu\, A|\cV)=\mu\,\fTB( A|\cV)$ ~for
$\mu\ge0$, but it is not a fully linear function of $A$; in fact, the map is
super-additive (Proposition \ref{pr:11}). Hence, excluding the trivial case of
$\cV$ being an invariant subspace of $\rho_A$, there is no proper PVM with
some orthogonal projector $P'$ such ~$\lambda = \Tr(P' \rho_A)$. ~Of course,
it is always possible to devise an ad-hoc PVM to do the job, and in
fact~(Proposition \ref{pr:9})
\be
\lambda = \Tr(Q \rho_A)
,
\ee
where $Q$ is the orthogonal projector onto the subspace
~$\cV':=\rho_A^{-1/2}\ran(\rho_A)\cap \cV$.\footnote{Also
  ~$\lambda = \Tr(\Pi \rho_A)$, from \eq{2.25}. $\Pi$ is a projector operator,
  but it is non-Hermitian in general, and it also depends on the state.} ~For
$\rho_A>0$, the same construction can be extended along the lines of
\eqs{3.34} and \nec{3.34a} with orthogonal projectors $Q_k$ onto
~$\cW'_k:=\rho_A^{-1/2}\cW_k$. Nevertheless, such a PVM is of limited use since
the $Q_k$ depend on the very state $\rho_A$ to be measured. Also, the
(unnormalized) post-measurement state $Q\rho_A Q$ has support on ~$\cV'$ so it
would not be directly $\rho_B$.

In any case, if $\lambda$ (or rather $\log(\lambda)$), is
regarded as a kind of entropy, it is only natural that it is not accessible as
the expectation value of a projector independent of the state, since this is
also the case of the von Neumann entropy.

\subsection{ Potential cryptographic application }

Let $\rho$ be a state with support contained in a subspace
$\cV\varsubsetneq\cH$, where the inequality can always be achieved by
adding ancillary states or otherwise enlarging $\cH$. Such state can be masked by
considering another state $\sigma$ in $\cH$ such that
~$\ran(\sigma)\cap\cV =0$, thus granting a disjoint support with
$\rho$. The state $\rho$ can be encrypted within a new state $\rho'$
through the mixing
\be
\rho' = \lambda \rho + (1-\lambda) \sigma,
\qquad
0 < \lambda < 1
\,,
\ee
for instance, using an ancillary qubit $A'$ and a standard
construction
\be
\rho' = \Tr_{A'} \big(
\lambda \rho \otimes \ketbra{0}
+ (1-\lambda) \sigma \otimes \ketbra{1} \big)
\,.
\ee
The state $\rho'$ can be public while $\cV$ is known only to the
receiver. Since $\rho$ and $\sigma$ have disjoint supports, $\rho'$
and $\cV$ univocally determine the state $\rho$ (as well as $\sigma$
and $\lambda$) through the Lebesgue decomposition.

An important observation is that the $\rho'$ should be explicitly
known by the receiver: it would not be sufficient to receive a copy
(or a few copies) of the state. This limitation could be circumvented
by choosing $\cV$ as one of the invariant subspaces of $\rho'$, since
in that case an ideal measurement of $P$ on $\rho'$ would directly
select $\rho$. However, if $\rho'$ is known by a eavesdropper, the
condition on $\cV$ of being an invariant subspace is too restrictive
and hence it gives away too much information on $\rho$. Allowing a
more general $\cV$ is much less informative while still uniquely
determines $\rho$.

\section{  Decomposition of a non-negative operator along a subspace }
\label{sec:1}

The basic statement of the Lebesgue decomposition of positive
operators is that expressed in the Theorem \ref{th:intro}.  In this
section the mathematical structure is developed in the simpler
scenario of finite-dimensional Hilbert spaces, to support the analysis
presented in the previous section. In addition to the decomposition itself,
further mathematical properties of the decomposition are derived; they
may or may not admit an extension to the infinite-dimensional case.

{\bf Some conventions and remarks:} all abstract vector spaces
considered in this work (in particular Hilbert spaces) are complex and
finite-dimensional, unless otherwise specified (hence, operators are
bounded, subspaces are closed, $\ran(A^{1/2})=\ran(A)$, etc). For
Hilbert spaces, the symbol $\oplus$ denotes the direct sum of
orthogonal subspaces, ~$\subseteq$ denotes subspace, and $\ominus$
denotes orthogonal complement. The symbol $\dplus$ denotes direct sum
(not necessarily orthogonal) and $I$ denotes the identity
operator. The acronym PVM stands for {\em projector valued measure},
that is, a set of non-zero operators $\{P_k\}_{k=1}^n$, such that
~$P_k^\dagger = P_k$, ~$P_k P_j = \delta_{kj}P_k$ ~and
~$\sum_{k=1}^n P_k=I$.  The scalar product is conjugate-linear on the
first argument.  For an orthogonal projector $P$ onto some subspace,
$P^\perp := I-P$. For a subspace $\cV \subseteq \cH$,
$\cV^\perp := \cH\ominus \cV$. The range of an operator always refers
to its image, $\ran(A)=A(\cH)$.

\subsection{  Main Theorem }
\label{sec:1a}

\medskip \medskip %
\newdefinition{df:1} {\em (Support and disjoint operators)} ~The image
$\ran(X)$ of a normal operator $X$ will be called its {\em support} (in the
sense of linear algebra).\footnote{Regarding $X$ as a function, its support is
  the complementary of $\ker(X)$ as a set, while $\ran(X)$ is the orthogonal
  complement as a subspace.}  Likewise, two normal operators $X$ and $Y$ will
be called {\em disjoint} when ~$\ran(X) \cap \ran(Y) = 0$.  \fin

\medskip \medskip %
\newproposition{pr:1} ~{\em Let $X$ and $Y$ be two Hermitian and disjoint
  operators in a Hilbert space $\cH$, ~then
\begin{itemize}

\item[1)] $\ran(X+Y) = \ran(X) + \ran(Y)$.

\item[2)] When ~$X+Y=I$, ~the set ~$\{X,Y\}$ ~is a PVM in ~$\cH$.

\item[3)] If ~$X + Y \ge 0$ ~then ~$X \ge 0$ ~and ~$Y \ge 0$. In addition, when
  ~$X + Y > 0$, ~$\ran(X) \dplus \ran(Y) = \cH$.

\end{itemize}
}

\medskip \proof
\begin{itemize}

\item[1)] Clearly ~$\ran(X+Y) \subseteq \ran(X) + \ran(Y)$. Let us show that
  $ (\ran(X) + \ran(Y) ) \ominus \ran(X+Y) = 0$: Since $X+Y$ is a normal
  operator, $\psi \perp \ran(X+Y)$ iff $\psi \in \ker(X+Y)$. In that case,
  $(X+Y)\psi =0 \implies X\psi = - Y \psi =0$ due to the following:
  $\ran(X) \cap \ran(Y) =0$, and $\psi \in \ker(X) \cap \ker(Y)$. Then
  $\psi \perp \ran(X) + \ran(Y)$; it follows that
  $\ran(X+Y) = \ran(X) + \ran(Y)$.
  
\item[2)] Since the Hermitian operators $X$ and $Y$ commute, they share a
  common spectral decomposition
\be
X = \sum_{k=1}^n x_k P_k, \qquad
Y = \sum_{k=1}^n y_k P_k,
\ee
where $\{P_k\}_{k=1}^n$ is a PVM and $x_k,y_k\in\R$. The condition
~$\ran(X) \cap \ran(Y) = 0$ ~requires that ~$\forall k ~x_k y_k = 0 $. The
condition $X+Y=I$ ~implies that ~$\forall k ~x_k + y_k = 1 $. Then, for each
$k$ either $x_k=1$ and $y_k=0$ or $x_k=0$ and $y_k=1$, hence the set $\{X,Y\}$
is a PVM.

\item[3)] Let $M:= X + Y$.  If ~$M=0$ ~it follows from clause 1 that
  ~$X=Y=0$. ~Otherwise we can restrict ourselves to the non-zero subspace
  $\cR := \ran(M)$. Within $\cR$ the operator $M$ is positive. Then
  \be I = M^{-1/2} X M^{-1/2} + M^{-1/2} Y M^{-1/2} := X' + Y' .
  \ee
  Clearly $X'$ and $Y'$ are Hermitian and disjoint; hence, they form a PVM in
  $\cR$. Then $X$ and $Y$ are positive operators in $\cR$ and non-negative
  in $\cH$. When $M>0$ in $\cH$, $\cH=\ran(M)= \ran(X)+\ran(Y)$.  \qed
  
\end{itemize}

The clauses 1 and 2 of the Proposition can be relaxed to require only $X$,
$Y$ and $X+Y$ to be normal operators.

The main theorem is as follows:

\medskip \medskip %
\newtheorem{th:1} ~{\em Let $A \ge 0$ be a non-negative operator in a Hilbert space
$\cH$, and $\cV \subseteq \cH$. Then $A$ admits a decomposition ~$A = B + C$,
~such that
\begin{itemize}
  \item[$i)$] $B$ and $C$ are Hermitian, and
  \item[$ii)$] ~$\ran(B) \subseteq \cV$ ~and ~$\ran(C) \cap \cV = 0$.
\end{itemize}
The decomposition is unique, and the operators $B$ and $C$ are non-negative.
}

\medskip%
Clearly this is a consequence of Theorem \ref{th:intro}, $\cH$ being
finite-dimensional.

\medskip%
\proof ~If $\cV=0$, the unique decomposition $B=0$, $A=C$ fulfills the
statement, and  similarly, if $\cV^\perp=0$ with $C=0$ and $A=B$. Then, we 
assume that $\cV$ and $\cV^\perp$ are non-vanishing.

Let $P$ denote the orthogonal projector onto $\cV$ and let us make the
decomposition $\cH = \cV \oplus \cV^\perp$ and then
$\cV^\perp = \cR \oplus \cK$, where $\cK$ is the kernel of $P^\perp A P^\perp$
within $\cV^\perp$. In block matrix form, with decomposition
$\cH = \cV \oplus \cR \oplus \cK$,
\be
A = \PM{ a & b^\dagger & h^\dagger \\ b & c & 0 \\ h & 0 & 0}
,
\ee
where, for instance ~$b: \cV \to \cR$, ~and the operator is extended as zero
when acting on $\cR$ or $\cK$, and similarly for $a$, $h$, and $c$. Let
$\psi \in \cV$ and $\chi \in \cK$, then
\be
0 \le (\psi + \chi)^\dagger A (\psi + \chi)
=\psi^\dagger a\psi + \psi^\dagger h^\dagger \chi + \chi^\dagger h \psi
.
\ee
The inequality for all $\psi$, $\chi$ requires $h=0$,
\be
A = \PM{ a & b^\dagger & 0 \\ b & c & 0 \\ 0 & 0 & 0}
.
\ee

When $\cR=0$, ~$A=a$, ~then $A=B$ and $C=0$ is the unique decomposition, and
the statement is fulfilled. In the following, we assume that $\cR$ is non-zero.
Then $c>0$ within $\cR$. Let us define the Hermitian operators $B$ and $C$ as
follows:
\be
B:=  \PM{ a - b^\dagger c^{-1} b  & 0 &0 \\ 0 & 0 & 0 \\ 0 & 0 & 0 },
\qquad
C:= \PM{ b^\dagger c^{-1} b & b^\dagger & 0 \\ b & c & 0  \\ 0 & 0 & 0  }
.
\ee
By construction ~$A=B+C$ ~and ~$\ran(B)\subseteq \cV$. ~The operator
~$a - d c^{-1} b$ ~is known as the Schur complement of the block $c$ in the
matrix $\PM{ a & d \\ b & c}$ \cite{Zhang:2005,Horn:2012,Bhatia:2007}.

Let us show that $B$ is non-negative. Let $\psi \in \cV$ and $\phi \in \cR$,
then
\be
0 \le (\psi + \phi)^\dagger A (\psi + \phi) =
\psi^\dagger a \psi + \psi^\dagger b^\dagger \phi
+
\phi^\dagger b \psi  + \phi^\dagger c \phi .
\ee
For given $\psi$, ~$\phi= - c^{-1} b \psi$ ~attains the minimum and yields
\be
\forall \psi \in \cV \quad
0 \le \psi^\dagger (a- b^\dagger c^{-1} b ) \psi,
\ee
implying that $B\ge 0$. Likewise, $C$ is non-negative, indeed,
\be
C = \PM{b^\dagger c^{-1/2} \\ c^{1/2} \\ 0} \PM{  c^{-1/2} b & c^{1/2} & 0}
:= f^\dagger f \ge 0
.
\ee
Here ~$f : \cH \to \cR$ ~and ~$f^\dagger : \cR \to \cH$.
The kernel and range of $C$ are readily computed:
\bes
\ker(C) &= \{
\PM{ \psi \\ -c^{-1}b \psi \\ \chi } ~|~ \psi \in \cV, ~ \chi \in \cK 
\,\}
,
\\
% \qquad
\ran(C) &= \{
\PM{  b^\dagger c^{-1} \phi \\ \phi  \\ 0 } ~|~ \phi \in \cR 
\,\}
.
\ees
It follows that ~$\ran(C) \cap \cV = 0$.

It remains to show that the decomposition ~$A = B + C$ ~is unique. Let
~$A = B' + C'$ ~be another decomposition, with $B'$ and $C'$ Hermitian,
~and ~$\ran(B')\subseteq \cV$ ~and ~$\ran(C') \cap \cV = 0$. ~Since $B$ and $B'$
vanish on $\cV^\perp$, it follows that $C'$ must follow the pattern
\be
C' = C + \PM{  d & 0 & 0 \\ 0 & 0 & 0  \\ 0 & 0 & 0 }
,
\ee
then 
\be
\forall \psi \in \cV \quad C' (\psi -c^{-1}b \psi ) =
 d \psi \in \ran(C') \cap \cV = 0
\ee
requires $d=0$, hence $C'=C$ and $B'=B$. \qed

\medskip%
In general, the spaces $\ran(B)$ and $\ran(C)$ need not be orthogonal.
 
\medskip%
From the inspection of the uniqueness part of the proof, it follows that the
same decomposition of $A \ge 0$ exists and is unique by requiring only $B$ and
$C$ to be normal operators such that ~$\ran(B)\subseteq \cV$ ~and
~$\ran(C) \cap \cV = 0$. Also, we remark that the weaker condition
requiring $B$ and $C$ to be merely disjoint would not guarantee uniqueness.

The uniqueness of the decomposition implies that, if $B'\neq B$ is Hermitian
and such that ~$\ran (B')\subseteq \cV$, ~necessarily
~$\ran(A-B')\cap \cV \neq 0$.

\medskip \medskip%
\newdefinition{df:2} ~Under the conditions of the Theorem \ref{th:1}, the
operator $B$ will be called the component of $A$ along $\cV$ as a subspace of
$\cH$, and ~$A=B+C$ ~the decomposition of $A$ along $\cV$ as a subspace of
$\cH$.\footnote{When the decomposition is written as $A=B+C$ it will be tacitly
  understood that the first summand $B$ is the component along the subspace.}
~For convenience, we will introduce the functions
\be
\begin{aligned}
\fB(A|\cV) &:= B, 
  &
    \fTB(A|\cV) &:= \Tr(B),
     \\
\fC(A|\cV) &:= C, 
  &
\fTC(A|\cV) &:= \Tr(C),
\\
\fP(A|\cV) &:= PAP, 
  &
\fTP(A|\cV) &:= \Tr(PA),
\ignore{
\\
\fV(A|\cV) &:= \ran(B), 
  &
\fW(A|\cV) &:= \ran(C),
}
\end{aligned}
\label{eq:2.16}
\ee
where $P$ denotes the orthogonal projector operator onto $\cV$. More
generally, we can define $\fB(A|\cH,\cV)$, etc, when the full space needs to
be specified.  \fin

\medskip \medskip%
\newcorollary{co:2} ~{\em For $A \ge 0$, let ~$A=B+C$ ~be the decomposition in
  $\cH$ along $\cV \subseteq \cH$. Then
\be
\ran(B) \dplus \ran(C) = \ran(A) \,.
\ee
}

\medskip %
\proof ~It follows from the Proposition \ref{pr:1} since $B$ and $C$ are
disjoint. \qed

%-------------------------------
\medskip \medskip%
Further information on the structure of the decomposition is obtained from the
following theorem:

\medskip \medskip%
\newtheorem{th:4} ~{\em For $A \ge 0$, let ~$A = B + C$ ~be its decomposition
along $\cV$ as a subspace of $\cH$, and let $\tcV:= \cV \cap \ran(A)$.
\begin{itemize}
\item[1)] $A = B + C$ ~is also the decomposition of $A$ along $\tcV$ as a
  subspace of\/ $\ran(A)$.
\item[2)] $\rank(B) + \rank(C) = \rank(A)$.
\item[3)] $\ran(B) = \tcV$.
\end{itemize}
}

\medskip%
\proof
\begin{itemize}
\item[1)] Because the Corollary \ref{co:2} ~$B$ and $C$ vanish on ~$\ker(A)$,
  ~thus $A$, $B$, and $C$ can be regarded as operators on $\ran(A)$.  ~The
  statement then follows from ~$\ran(B) \subseteq \tcV$ ~and
  ~$\ran(C) \cap \tcV = 0$, ~plus uniqueness of the decomposition.

\item[2)] The statement follows from the Corollary \ref{co:2}.

\item[3)]  The relations
~$\ran(B) \subseteq \tcV$, ~and ~$\ran(C) \cap \tcV =0$  imply that
~$\rank(B) \le \dim \tcV$ ~and ~$\rank(C) + \dim \tcV \le \rank(A)$. Hence
\be
\rank(B) + \rank(C) \le \rank(A)
.
\ee
Reaching the equal sign requires ~$\rank(B) = \dim \tcV$ and in turn
~$\ran(B) = \tcV$. \qed
  
\end{itemize}

\medskip \medskip %
The Theorem implies that the decomposition effectively occurs within
$\ran(A)$ and it is unchanged if $\cV$ is replaced by ~$\cV \cap \ran(A)$.
  
\medskip \medskip %
\newcorollary{co:4} ~{\em The decomposition of $A \ge 0$ along $\cV$ in $\cH$ reduces
to that of ~$A > 0 $ ~within ~$\ran(A)$ ~along ~$\cV \cap \ran(A)$.
}

\medskip \medskip %
The corollary implies that the decomposition of a non-negative operator $A$ on
$\cH$ can always be reduced to that of the positive operator $A$ on the subspace
~$\ran(A)$.

%-------------------------------
\medskip \medskip %
Some immediate consequences of the Theorems \ref{th:1} and \ref{th:4} are as
follows:

\medskip \medskip %
\newproposition{pr:10} ~{\em For $A \ge 0$, let ~$A=B+C$ ~be its decomposition
along in $\cV$ as a subspace of $\cH$. Then

\begin{itemize}

\item[1)] $\ran(A) \subseteq \cV  \iff C=0$.

\item[2)] $\ran(A) \cap \cV = 0 \iff B=0$.
  
\item[3)] If ~$\cH \subseteq \cH'$, ~$A=B+C$ ~is also the decomposition of $A$
  along $\cV$ as a subspace of $\cH'$. $A$, $B$, and $C$ are extended as
  operators on $\cH'$ ~which vanish on ~$\cH' \ominus \cH$.

\item[4)] Let $\cW:=\ran(C)$, then ~$A=C+B$ ~is the decomposition of $A$ along
  $\cW$ as a subspace of $\cH$.

\item[5)] When $\cV$ and $\cV^\perp$ are invariant subspaces of $A$,
\be
\fB(A| \cV ) = \fP(A|\cV), \qquad
\fC(A| \cV ) = \fP(A|\cV^\perp).
\ee

\item[6)] For $\mu \ge 0$,
\bes
\fB(\mu \, A| \cV ) &= \mu \, \fB(A|\cV), \\
\fC(\mu \, A| \cV ) &= \mu \, \fC(A|\cV).
\ees

\item[7)] Let $B'$ be a Hermitian operator such that ~$A+B'\ge 0$ ~and ~$\ran(B') \subseteq \cV$, ~then
\bes
\fB(A+B' |\cV) &= \fB(A |\cV) + B'
\,,
\\
\fC(A+B' |\cV) &= \fC(A |\cV) 
\,.
\ees

\item[8)] Let $U$ be a unitary operator in $\cH$, then
\be
\fB( U A U^\dagger | U\cV) = U \fB(A | \cV) U^\dagger
.
\ee

\end{itemize}
}

\medskip %
\proof

1) and 2) follow from ~$A=A+0$ ~and ~$A=0+A$, ~being respectively valid
decompositions in each case.
  
  3) follows from the uniqueness of the decomposition.

  4) The exchange of roles between and $B$ and $C$ follows from the uniqueness
  of the decomposition since $B$ and $C$ are Hermitian,
  ~$\ran(C)\subseteq \cW$ ~and ~$\ran(B) \cap \cW =0$.

  5) When $\cV$ is an invariant subspace ~$A = P A P + P^\perp A P^\perp
  $. Such decomposition fulfills the conditions of the Theorem \ref{th:1} and
  the statement follows from uniqueness.

  6) follows from the uniqueness of the decomposition.
  
  7) It follows from uniqueness, since $A+B'=(B+B')+C$, ~fulfills the
  conditions of the unique decomposition of $A+B'$ along $\cV$. The statement
  is also immediate from the structure of the Schur complement, namely,
  ~$B=a - b^\dag c^{-1} b$.

  8) The decomposition ~$ U A U^\dagger = U B U^\dagger + U C U^\dagger $
  ~fulfills the conditions of the Theorem \ref{th:1} and the statement follows
  from uniqueness.

  \qed

  \medskip%
  It can be noted that in the clause 7 nothing is assumed about the positivity
  of $B'$, still whenever $\ran(B')\subseteq\cV$, ~$A+B'\ge 0$ implies
  ~$B+B' \ge 0$. This corollary is equivalent to the Lemma \ref{lm:1}
  below.

\subsection{  Concavity of ~$\fB(A|\cV)$  }
\label{sec:1b}

\medskip \medskip%
\newlemma{lm:1a} ~{\em Let ~$A$ ~and ~$B$ ~be two non-negative operators such that
~$B\le A$, ~then ~$\ran(B) \subseteq \ran(A)$.
}

\medskip %
\proof
~$0 \le B \le A$ ~implies ~$\ker(B) \supseteq \ker(A)$, ~and thus
 ~$\ran(B) \subseteq \ran(A)$.
\qed

\medskip \medskip%
Note that ~$\ran(B) \subseteq \ran(A)$ ~is not guaranteed by $B\le A$ without
$0\le B$.

\medskip \medskip%
\newlemma{lm:1} ~{\em Let ~$\cV \subseteq \cH$, ~and ~$A\ge0$ in $\cH$ ~and let
~$A=B+C$ ~be its decomposition along $\cV$.  Let $B'$ be an Hermitian operator
such that ~$\ran(B') \subseteq \cV$ ~and ~$B'\le A$, ~then ~$B' \le B$. ~When
~$B'\ge 0$,  ~$\ran(B') \subseteq \ran(B)$.
}

\medskip %
\proof
\be
A = B+C \ge B'
\implies M:= (B-B') + C \ge 0
.
\ee
$\ran(B-B') \subseteq \cV$ ~and ~$\ran(C) \cap \cV=0$ ~thus ~$B-B'$ and $C$
are disjoint. Then clause 3 of the Proposition \ref{pr:1} applies and
~$B-B'\ge 0$.  ~The implication ~$B'\ge 0\implies \ran(B') \subseteq \ran(B)$
~follows from the Lemma \ref{lm:1a}. \qed

\medskip %
The Lemma implies that
\bes
\fB(A|\cV) &= \max\{ B' ~|~ B'\le A, ~~\ran(B') \subseteq \cV \}
,
\\
\fC(A|\cV) &= \min\{ C' ~|~ C'\ge 0, ~~\ran(A-C') \subseteq \cV \}
,
\label{eq:2.2}
\ees
thereby providing another way of uniquely characterizing the components $B$
and $C$.

For given $A \ge 0 $ and $\cV$ one could also consider the set of operators
defined by
\be
\{ C' ~|~ C' \le A, ~\ran(C') \cap \cV = 0 \},
\ee
however the condition ~$\ran(C')\cap\cV = 0$ ~is not sufficiently restrictive,
so in general, this set has no maximal element and does not single out
$C=\fC(A|\cV)$ ~as a distinguished element.

\medskip \medskip%
Let us highlight some further properties of the decomposition.

\medskip %
\newproposition{pr:3} ~{\em For $A \ge 0$, let ~$A=B+C$ ~be its decomposition in
$\cH$ along $\cV \subseteq \cH$.

\begin{itemize}

\item[1)] If ~$\cV \subseteq \cV'$ ~and ~$A = B' + C'$ ~is the decomposition
  along $\cV'$, then ~$B \le B'$.
  
\item[2)] If ~$A \le A'$ ~and ~$A' = B' + C'$ ~is the decomposition of $A'$
  along $\cV$, then ~$B \le B'$.
  
\item[3)] Let ~$\cV_1\subseteq\cV$, ~and let ~$A = B_1 + C_1$ ~and
  ~$B=B_1' + C_1'$ ~be the decompositions along ~$\cV_1$, ~then
  ~$B_1' \le B_1 \le B$.
  
\end{itemize}
}

\medskip %
\proof ~The statements are a straightforward consequence of the Lemma
\ref{lm:1}.  \qed

\medskip \medskip%
\newproposition{pr:11} ~{\em The map ~$A \to \fB(A|\cV)$ ~is super-additive,
  while ~$A \to \fC(A|\cV)$ ~is sub-additive, that is,
\bes
\fB(A_1+A_2|\cV) &\ge   \fB(A_1|\cV) + \fB(A_2|\cV)
,
\\
\fC(A_1+A_2|\cV) &\le   \fC(A_1|\cV) + \fC(A_2|\cV)
.
\ees
}

\medskip %
\proof ~If ~$A_1 = B_1 + C_1$, ~$A_2 = B_2 + C_2$, ~and $A_1+A_2 = B+C$, ~are
the decompositions along $\cV$, ~$B_1+B_2\le A_1+A_2$ and
$\ran(B_1+B_2)\subseteq\cV$, thus by the Lemma \ref{lm:1} ~$B\ge B_1+B_2$.
~The sub-additivity of $\fC(A|\cV)$ ~follows from $\fC(A|\cV)=A - \fB(A|\cV)$.
\qed

\medskip \medskip%
\newtheorem{th:2} ~{\em For given $\cV\subseteq \cH$, the maps ~$\fB(A|\cV)$ ~and
~$\fTB(A|\cV)$ ~are concave functions with respect to their dependence on
$A \ge 0$. Likewise the maps  ~$\fC(A|\cV)$ ~and
~$\fTC(A|\cV)$ ~are convex.
}

\medskip%
\proof ~The concavity of the map ~$A \to \fB(A|\cV)$,
\bes
\fB(t A_1&+(1-t)A_2|\cV) \ge   t \fB(A_1|\cV) + (1-t)\fB(A_2|\cV),
\\ & 0\le t \le 1
\ees
is an immediate consequence of super-additivity plus the positive homogeneity
property ~$\fB(\mu\, A|\cV) = \mu \,\fB(A|\cV)$ ~for ~$\mu\ge 0$ ~(clause 6 of
the Proposition \ref{pr:10}). ~And similarly for the convexity of
~$A \to \fC(A|\cV)$. ~The traces inherit these properties.  \qed

\medskip \medskip %
The concavity of the map ~$A \to \fTB(A|\cV)$ ~implies its continuity within
the open set of positive operators. The map is not continuous on the closed
set of non-negative operators. For instance, if $A$ is the orthogonal
projector onto the one-dimensional subspace spanned by a unit vector $\psi$,
it is readily found that ~$\fTB(A|\cV) = 0$ ~whenever ~$\psi \not\in \cV$ ~but
~$\fTB(A|\cV) = 1$ ~when ~$\psi\in \cV$. The continuity of
~$A \to \fTB(A|\cV)$ for $A>0$ also follows from the Proposition \ref{pr:7}
below. The continuity of the map ~$\cV \to \fTB(A|\cV)$ is discussed in the
Appendix \ref{app:a}

\medskip \medskip%
\newcorollary{co:5} ~{\em The map ~$A \to \log ( \fTB(A|\cV))$ ~is concave.}

\medskip%
\proof ~The statement follows from concavity of $A \to \fTB(A|\cV)$ and the
fact that the logarithm function is both increasing and concave. \qed

\subsection{  Alternative forms of the decomposition  }
\label{sec:1c}

\medskip \medskip %
The proof of the Theorem \ref{th:1} provides a concrete construction of the
components in $A=B+C$, namely, through the Schur complement.  The same
decomposition can be obtained from alternative constructions.

\medskip \medskip%
\newdefinition{df:3} For a normal operator $X$ and $n\in\Z$, we use the
notation $X^n$ for the normal operator corresponding to $X^n$ on $\ran(X)$ and
zero on $\ker(X)$.  With this definition ~$\ran(X^n)=\ran(X)$,
~$X^n X^m=X^{n+m}$, ~$(X^n)^m=X^{nm}$, ~and $X^0$ is the orthogonal projector
onto ~$\ran(X)$. For $A\ge0$ and $\kappa\in\C$, the operator $A^\kappa$ denotes
$A^\kappa$ on $\ran(A)$ and zero on $\ker(A)$.  \fin

\subsubsection{  Alternative form from projection  }
\label{sec:1c1}

\medskip \medskip%
\newproposition{pr:9} ~{\em Let $A=B+C$ be the decomposition of $A\ge 0$ along
  the subspace $\cV\subseteq\cH$. Let $Q$ be the orthogonal projector operator
  onto the subspace ~$A^{-1/2} \tcV$ ~where ~$\tcV:=\ran(A)\cap\cV$. ~Then
\bes
&1) \qquad B = A^{1/2} Q A^{1/2}, \qquad
C = A^{1/2} (I-Q)A^{1/2}
      . 
      \\
&2) \qquad \Tr(B) = \Tr(QA)
       .
       \\
&3) \qquad BA^{-1}B = B ,\qquad
BA^{-1}C = 0
\,.
 \label{eq:2.22}
\ees
}

\medskip%
\proof 
\begin{itemize}
\item[1)] Let
  \be
  B' := A^{1/2} Q A^{1/2}, \quad
  C' := A^{1/2} (I-Q)A^{1/2}
  .
  \ee
  By construction $B'$ and $C'$ are Hermitian and ~$A=B'+C'$.  Also
  ~$\ran(B') =\tcV \subseteq \cV$. On the other hand, since
  $\ran(C')\subseteq\ran(A)$, ~$\ran(C')\cap \cV \subseteq\tcV$, but $C'$ is
  disjoint from $B'$, thus ~$\ran(C')\cap \cV=0$. Then ~$B'=B$ ~and ~$C'=C$
  ~by uniqueness of the decomposition.
  
\item[2)] It is a consequence of the previous clause.

\item[3)] The relations follow from the property ~$Q^2=Q$.
  
\end{itemize}
\qed

\medskip \medskip%
Regarding the second clause, it should be noted that the operator $Q$ depends
on $A$, as well as on $\cV$.

\medskip \medskip%
For $A>0$, the spaces $\cV=\ran(B)$ and $\cW:=\ran(C)$ are not orthogonal, in
general. Nevertheless, they can be regarded as orthogonal by modifying the
scalar product, namely: if $\psi,\phi\in\cH$ and $\psi^\dag \phi$ denotes the
(standard) scalar product defined in the Hilbert space $\cH$, the subspaces
$\cV$ and $\cW$ are orthogonal with respect to the new scalar product
\be
\psi^\ddag \phi := \psi^\dag A^{-1} \phi
,
\label{eq:2.25a}
\ee
as can readily be verified. If $\Pi$ and $I-\Pi$ denote the $\ddag$-orthogonal
projectors onto $\cV$ and $\cW$, respectively, $B$ and $C$ are recovered as
follows:
\be
B = \Pi A, \qquad
C= (I-\Pi)A
.
\label{eq:2.25}
\ee

More generally, for $A>0$, let
\be
\cH = {\mathop{\Oplus}\limits_{k=1}^n} \!{}^\ddag \,\cW_k
\,,
\qquad
I = \sum_{k=1}^n \Pi_k 
\,,
\ee
where the subspaces $\cW_k$ are $\ddag$-orthogonal, and the $\Pi_k$ are the
corresponding $\ddag$-orthogonal projectors. Then $A$ gets decomposed into
non-negative operators as follows:
\be
A = \sum_{k=1}^n C_k, \qquad
C_k = \Pi_k A
\,.
\ee
By construction ~$\ran(C_k) \subseteq \cW_k$ ~(in fact they are equal), ~and
~$\ran(C_k)\cap \cW_j = 0$ ~when ~$j\neq k$. The $C_k$ are non-negative (with
respect to the standard scalar product),\footnote{The non-negative character of
  an operator depends on the scalar product. $A>0$ under both definitions.
  $\Pi_k$ is $\ddagger$-non-negative, being a $\ddagger$-orthogonal
  projector. \Eq{2.31} then implies that $C_k$ is $\dagger$-non-negative (and
  thus $\dagger$-Hermitian).} ~indeed
\be
\psi^\dagger C_k  \psi =
\psi^\dagger \Pi_k A \psi =
\psi^\ddagger A \Pi_k A \psi \ge 0
\,,
\label{eq:2.31}
\ee
due to ~$A=A^\ddag$~ and ~$\Pi_k= \Pi_k^\ddag $.

By the uniqueness of the decomposition, each $C_k$ is the
component of $A$ along $\cW_k$. ~It should be observed that, although the
$C_k$ have $\ddag$-orthogonal supports, ~$A = \sum_{k=1}^n C_k$ ~is not a
decomposition of $A$ into invariant subspaces since the $C_k$ are not
$\ddag$-Hermitian.

An explicit instance of the above construction follows from having a chain of
inclusions
\be
\cH \supseteq \cV_1 \supseteq \cdots \supseteq \cV_{n-1} \supseteq \cV_n=0
\,.
\ee
Then ~$A>0$ ~can be decomposed recursively as
\bes
B_0 &:=A, \\
B_{k-1} &=: B_k + C_k \quad\text{(along $\cV_k$)},
\\
\cW_k&:=\ran(C_k),
\qquad
k=1,\ldots, n
\,.
\ees
At the end of the recursion, ~$B_n=0$, ~$C_n=B_{n-1}$ ~and ~$\cW_n=\cV_{n-1}$.
This gives
\be
A = B_1 + C_1
=
(B_2+C_2) + C_1 = \cdots
= C_n + \cdots + C_1
\,.
\ee
Here, the subspaces $\cW_k$, for $k=1,\ldots,n-1$, are generated by the
decomposition and not chosen a priori.

\subsubsection{  Alternative form from $A^{-1}$  }
\label{sec:1c2}

\medskip \medskip%
\newproposition{pr:7} ~{\em Let $A\ge0$ on $\cH$ and $A=B+C$ be its
  decomposition along $\cV\subseteq \cH$, then
\be
B^{-1} = \tP A^{-1} \tP
\,,
\label{eq:2.32}
\ee
where $\tP$ is the orthogonal projector operator onto the subspace
~$\ran(A) \cap \cV$.
}

\medskip%
\proof ~Due to the Corollary \ref{co:4}, the operator $B$ is the component of
$A$ along $\tcV:= \ran(A) \cap \cV$ as a subspace of $\ran(A)$. The latter can
be decomposed as
\be
\ran(A) = \tcV \oplus \tcV^\perp
,
\ee
where ~$\tcV^\perp:= \ran(A)\ominus \tcV$. Then, the corresponding block
decompositions of $A$ and $A^{-1}$ take the form
\be
A = \PM{ a & b^\dagger  \\ b & c   },
\qquad
A^{-1} = \PM{ \alpha & \beta^\dagger   \\ \beta & \gamma }
.
\ee
Positivity of $A$ on $\ran(A)$ ensures that ~$c>0$ and $B=a-b^\dag c^{-1} b$,
from the construction in the Theorem \ref{th:1}. ~The condition ~$AA^{-1}= I$
~on $\ran(A)$ implies
\be
I= a\alpha+b^\dagger \beta
,
\qquad
0 = b \alpha + c \beta
.
\ee
Thus ~$\beta = -c^{-1} b \alpha$ ~and so
~$I= (a-b^\dag c^{-1} b )\alpha = B\alpha$. ~As a consequence
$\alpha = B^{-1}$ ~as an operator on ~$\tcV$.  \qed

\medskip \medskip%
Note that the simpler relation $B^{-1} = P A^{-1} P$, where $P$ is the
orthogonal projector onto $\cV$, holds when $A>0$ but not in general.

\medskip \medskip%
In the particular case of $\cV$ being one-dimensional, spanned by a 
non-zero vector ~$\psi \in \ran(A)$, ~\eq{2.32} takes the form
\be
B = \frac{\ketbra{\psi}}{\esp{\psi |A^{-1} |\psi}}
.
\label{eq:3.29}
\ee

\medskip \medskip%
For given $A\ge 0$, the map $\fTB(A|\cV)$, as a function of $\cV$, contains
sufficient information to completely characterize the operator $A$. That is,
for ~$A_1,A_2\ge 0$,
\be
\forall \cV ~~\fTB(A_1|\cV) = \fTB(A_2|\cV)
~\implies ~A_1=A_2
,
\ee
and in fact, the subset of one-dimensional subspaces suffices:

\medskip \medskip%
\newproposition{pr:5} ~{\em The map $\cV \to \fTB(A | \cV)$ restricted to the
  set ~$\{\cV \,|\, \dim \cV=1\}$ ~uniquely determines the non-negative
  operator $A$.  }

\medskip%
\proof ~It follows from \nec{3.29}: The $\psi$ with a non-vanishing value of
$\fTB(A | \mspan(\psi))$ reveal the subspace $\ran(A)$ and there, the operator
$A^{-1}$ gets fully characterized by the values of
~$\esp{\psi |A^{-1} |\psi}/\esp{\psi|\psi} = \fTB(A | \mspan(\psi))^{-1}$.  \qed

\medskip \medskip%
A further inequality related to the direct product of spaces:

\medskip \medskip%
\newproposition{pr:12} ~{\em
  Let $A_1 \ge 0$ in $\cH_1$ and $A_2 \ge 0$ in $\cH_2$,
and let $\cV_1\subseteq\cH_1$ and  $\cV_2\subseteq\cH_2$. Then
\bes
\fB(A_1\otimes A_2|\cV_1\otimes\cV_2) &\ge \fB(A_1|\cV_1) \otimes
\fB(A_2|\cV_2)
,
\\
\fTB(A_1\otimes A_2|\cV_1\otimes\cV_2) &\ge \fTB(A_1|\cV_1) \, \fTB(A_2|\cV_2) 
\ees
}

\medskip%
\proof ~Let $A:=A_1\otimes A_2$ and $\cV := \cV_1\otimes \cV_2$.  The tensor
product of two non-negative operators is non-negative, hence $A \ge 0$. Let
$A_1=B_1+C_1$, $A_2=B_2+C_2$, and $A=B+C$ be the decompositions along $\cV_1$,
$\cV_2$ and $\cV$, respectively. $A \ge B_1\otimes B_2$ because
$A= B_1\otimes B_2 + B_1\otimes C_2 + C_1\otimes B_2 + C_1\otimes C_2$ and
each summand is non-negative.  Also
~$\ran(B_1 \otimes B_2)= \ran(B_1) \otimes \ran( B_2) \subseteq \cV$. Then,
from the Lemma \ref{lm:1} ~$B \ge B_1 \otimes B_2$.  The second equality
follows from taking the trace in the first one.  \qed

\medskip \medskip%
The relation $B=B_1\otimes B_2$ holds when $\cV_1$ and $\cV_2$ are invariant
subspaces of $A_1$ and $A_2$, respectively; in general the equality fails
because the property $\ran(A-B_1\otimes B_2)\cap\cV=0$ does not hold.

\subsection{  Trace inequalities }
\label{sec:1d}

\medskip \medskip%
\newproposition{pr:4} ~{\em Let $d=\dim\cH$, and ~let
~$a_1 \le a_2 \le \cdots \le a_d$ be the eigenvalues of\/ $A \ge 0$. Then
\be
T_- \le \fTB(A|\cV) \le T_+
\,,
\ee
where 
\be
T_-:= \sum_{k=1}^n a_k,
\qquad
T_+:= \sum_{k=d-n+1}^d a_k
\,,
\ee
and ~$n:= \dim(\cV \cap \ran(A))$.
}

\medskip%
\proof ~Let $Q$ be the orthogonal projector onto ~$\tcV':=A^{-1/2} \tcV$
~where ~$\tcV:=\ran(A)\cap\cV$. Then ~$\rank(Q) =n$ ~since $A$ is invertible
on ~$\ran(\tcV)$. From the Rayleigh-Ritz variational principle
\cite{Galindo1990quantum}, the $n$ lowest/highest eigenvalues of $A$ are
lower/upper bounds of those of $QAQ$, regarded as an operator on $\tcV'$. The
statement then follows from \nec{2.22} which implies ~$\fTB(A|\cV)
=\Tr(QAQ)$. \qed

\medskip \medskip%
For a given value of $n$, the boundary values $T_\pm$ are attained by the
spaces $\cV_{\pm}$ spanned by the $n$ eigenvectors of $A$ with the
highest/lowest eigenvalues, as follows from clause 5 of the Proposition
\ref{pr:10}

\medskip \medskip%
\newtheorem{th:3} ~{\em For ~$A\ge 0$ in $\cH$, let
  ~$\cV=\Oplus_{k=1}^n \cV_k$, with orthogonal projectors
  $P= \sum_{k=1}^n P_k$, and let ~$A=B+C=B_k+C_k$ ~be the
  decompositions of $A$ along $\cV$ and $\cV_k$ respectively. Then
\begin{itemize}

\item[1)] $\ds  P_k B P_k \ge B_k$\,.
  
\item[2)] $\ds \Tr(P_k B) \ge \Tr(B_k)$, ~and ~$\Tr(P_k B) = \Tr(B_k)$ ~iff
  ~$[P_k,B]=0$\,.

\item[3)] $\ds \Tr(B) \ge \sum_{k=1}^n \Tr(B_k)$, ~and
  ~$\ds \Tr(B) = \sum_{k=1}^n \Tr(B_k)$ ~iff
  ~$\forall k ~~[P_k,B]=0$\,.

\item[4)] $\ds \Tr(B^{-1}) \ge \sum_{k=1}^n \Tr(B_k^{-1})$. A sufficient
  condition for equality to hold is ~$\cV\subseteq \ran(A)$\,.

\end{itemize}
}

\medskip %
\proof ~

\begin{itemize}

\item[1)] Since ~$\cV_k\subseteq\cV$ ~the first clause of the Proposition
  \ref{pr:3} implies ~$B \ge B_k$. Then
\be
P_k B P_k \ge P_k B_k P_k = B_k 
.
\ee

\item[2)] $\ds \Tr(P_k B) \ge \Tr(B_k)$ ~follows from the previous inequality
  and the cyclic property of the trace.

  The property ~$P_k B P_k \ge B_k$ implies the following
  equivalence:\footnote{If $A\ge B$ and $\Tr(A)\ge \Tr(B)$, then $\Tr(A-B)\ge0$ for
  $A-B\ge0$, and this implies $A-B=0$.}
\be
\Tr(P_k B) = \Tr(B_k) \iff P_k B P_k = B_k
,
\ee
thus, the second part of the statement is equivalent to
\be
 P_k B P_k = B_k \iff [P_k,B] = 0
\,.
\ee

Let us assume that ~$P_k BP_k = B_k$, then also $B \ge P_k BP_k$, and this in
turn requires $[P_k,B] = 0$. Indeed, taking the matrix element with a state
~$\xi \ket{\psi}+\ket{\phi}$ with $\psi\in\cV_k$ ~and ~$\phi\in\cV_k^\perp$,
\be
\xi^*\esp{\psi|B|\phi} + \xi \esp{\phi|B|\psi} + \esp{\phi|B|\phi}
\ge 0
\ee
for all $\xi\in\C$ requires ~$\esp{\psi|B|\phi}=0$.

Conversely, let us assume $[P_k,B]=0$. If ~$B=B'_k+C_k'$ ~is the decomposition
of $B$ along $\cV_k$, which is an invariant subspace of $B$, it follows that
$C_k'$ lies on ~$\cV_k^\perp$ ~and ~$B_k' = P_k B P_k$, thus,
\be
B_k' \ge B_k \,.
\ee
Moreover, the Proposition \ref{pr:3}, clause 3, implies that
\be
B_k' \le B_k \,.
\ee
It follows that $P_k B P_k = B_k' = B_k$.

\item[3)]

  The first part follows from $\Tr(P_k B) \ge \Tr(B_k)$ using ~$\sum_k P_k = P$
  ~and ~$PB=B$. The second part is also an immediate consequence of the
  previous clause.

\item[4)] Let ~$\tcV:=\ran(A)\cap \cV$ ~and ~$\tcV_k:=\ran(A)\cap \cV_k$, and
  let ~$\tP$ and $\tP_k$ ~be the corresponding orthogonal projector operators.
  From the Proposition \ref{pr:7} it follows that
\be
B^{-1} = \tP A^{-1} \tP ,\qquad
B_k^{-1} = \tP_k A^{-1} \tP_k .
\ee
Clearly, ~$\tcV_k \subseteq \tcV $, ~thus ~$\tP_k \tP  = \tP_k$
and so
\be
B_k^{-1} = \tP_k B^{-1} \tP_k .
\ee
In addition, the $\tcV_k$ are orthogonal among them, and
~$\ds \Oplus_{k=1}^n \tcV_k \subseteq \tcV \subseteq \cV $, ~hence
~$\sum_{k=1}^n \tP_k \le \tP \le P$, ~therefore
\be
\sum_{k=1}^n \Tr(B_k^{-1})  =
\sum_{k=1}^n \Tr(\tP_k B^{-1})  \le
\Tr( B^{-1})
.
\ee
When ~$\cV \subseteq \ran(A)$, ~$\tcV=\cV$ and $\tcV_k=\cV_k$, thus
~$ \sum_{k=1}^n \tP_k = \tP = P$.

\end{itemize}
\qed

\medskip \medskip%
According to the third clause of the Theorem
\be
\sum_{k=1}^n \fTB( A | \cV_k) \le
\fTB( A | \Oplus_{k=1}^n \cV_k )
,
\label{eq:2.44}
\ee
and, in this sense, the map $\cV \to \fTB( A | \cV)$ is super-additive.

\medskip \medskip%
Note that the inequalities for the traces do not extend to the operators
themselves. That is, a relation of the type
\be
B \ge \sum_{k=1}^n B_k 
,
\ee
is not valid general.

\medskip \medskip %
\newcorollary{co:1} ~{\em For ~$A \ge 0$ in $\cH$:
  
\begin{itemize}
  
\item[1)] Let $B$ be the component of $A$ along $\cV\subseteq\cH$ and $P$ the
  orthogonal projector onto $\cV$, then
  \be
  PAP \ge B \quad\text{and}\quad \Tr(P A) \ge \Tr(B),
  \ee
  and ~$\Tr(P A) = \Tr(B)$ ~iff
  ~$[P,A]=0$\,.

\item[2)] Let  $A>0$, 
  ~$\cH = \Oplus_{k=1}^n \cV_k$, with orthogonal projectors $P_k$, and let
  ~$A=B_k+C_k$ ~be the decompositions of $A$ along $\cV_k$. Then
  \be
  \Tr(A^{-1}) = \sum_{k=1}^n \Tr(B_k^{-1})
  \,.
  \ee

\end{itemize}
}

\medskip %
\proof ~The statements are a special case of clauses 1, 2, and 4 of Theorem
\ref{th:3}, obtained by taking $\cV$ and $B$ there to be $\cH$ and $A$ here,
respectively. In the second clause equality holds because $\ran(A)= \cH$ when
$A>0$. \qed

\medskip %
A further relevant proposition:

\medskip \medskip %
\newproposition{pr:13} ~{\em
  Let $A>0$ and $\cV\subseteq\cH$, and let
  \be
  A= B+ C = B^\perp + C^\perp
  \ee
  be the decompositions of $A$ along $\cV$ and $\cV^\perp$, respectively. Let
  the operator $D$ be defined by
  \be
  A = B + B^\perp + D
  .
  \label{eq:2.56a}
  \ee
  Then $\Tr(D)\ge 0$. Furthermore ~$D\ge 0 $ ~iff ~$\cV$, $\cV^\perp$ ~are
  invariant subspaces of $A$, and in this case $D=0$.}

\medskip %
\proof ~
From the Corollary \ref{co:1}
\be
\Tr(B) \le \Tr(PA),\qquad
\Tr(B^\perp) \le \Tr(P^\perp A)
.
\ee
Then ~$P + P^\perp = I$ ~implies ~$\Tr(B) + \Tr(B^\perp) \le \Tr(A)$, ~hence
~$\Tr(D)\ge 0$.

Let $Q$ and $Q_\perp$ be orthogonal projectors onto $\cV':=A^{-1/2}\cV$ and
$\cV'_\perp:=A^{-1/2}\cV^\perp$. They fulfill
\be
\rank(Q)+\rank(Q_\perp)= \dim\cH
.
\label{eq:2.56}
\ee
Using the Proposition \ref{pr:9}, the \eq{2.56a} becomes
\be
I = Q + Q_\perp  + A^{-1/2} D A^{-1/2}
\label{eq:2.56b}
.
\ee
If the two subspaces $\cV'$ and $\cV'_\perp$ are not orthogonal, then there
exists a vector $\psi\in\cV'$ such that $\esp{\psi|Q_\perp|\psi}>0$, while
$\esp{\psi|I|\psi}= \esp{\psi|Q|\psi}$, thus
~$\esp{\psi|A^{-1/2} D A^{-1/2}|\psi}<0$, and the operator $D$ is not
non-negative. When the two subspaces are orthogonal, \eq{2.56} implies
~$I = Q + Q_\perp$ ~and so ~$D=0$. However, the orthogonality requires
~$PA^{-1}P^\perp=0$ which in turn requires $[P,A^{-1}]=0$, hence $\cV$ is an
invariant subspace of $A$.  \qed

\medskip \medskip%
It can be noted that $D$ can be non-negative and non-zero if $A\ge0$ is
permitted. For instance, when $A$ has support on a subspace
$\cW$ such that ~$\cV\cap\cW= \cV^\perp \cap\cW= 0$, ~$B=B^\perp=0$ and $D=A$.

\section{ Summary and conclusions
  \label{sec:d}}

With an application to quantum information in mind, we have addressed
the mathematical problem of decomposing a non-negative operator
$A=B+C$ along a given subspace $\cV\subseteq\cH$ as a sum of two
non-negative operators, $B$ with support within $\cV$, and $C$ with
support disjoint from $\cV$. Such Lebesgue decomposition always exists
and is unique, and the same decomposition is obtained by requiring the
operator $B$ to be maximal under the constraint $B\le A$
\cite{Anderson:1975}. The decomposition then translates to quantum
systems in the form $\rho_A = \lambda \rho_B + (1-\lambda ) \rho_C$,
so that $\rho_A$ is expressed as a mixture of two quantum states
$\rho_B$ and $\rho_C$ localized inside and outside $\cV$,
respectively.

Section \ref{sec:2} discusses implications of the Lebesgue
decomposition for quantum states and quantum information. First, we
emphasize the relation between the support of a quantum state
$ \rho_A$ and the pure states whose mixture produces such a
state. Then, the formulas and relations derived in the previous
section are brought to the setting of quantum
information. Particularly interesting is the interpretation of
$\lambda$ as a probability, namely, the probability for a quantum
system in the state $\rho_A$ to be found completely included within
the subspace $\cV$. Such interpretation is sustained by the properties
of the maps $A\to B$ and $\cV \to B$ exposed before. The inequality
$\lambda\le p$ relating the weight $\lambda$ with the standard
quantum-mechanical overlap probability $p=\Tr(P\rho_A)$ is
discussed. It is shown that the sum of the inclusion probabilities
within $\cV$ and within $\cV^\perp$, $\lambda+\lambda^\perp$, is
bounded by one. This property was to be expected a priori,
consistently with the interpretation of $\lambda$ and
$\lambda^\perp$. Nevertheless, in general, there is a deficiency
$1-\lambda-\lambda^\perp \ge0$ that cannot be attributed to a true
quantum state $\rho_D$; in fact, $\Tr(\rho_D)=1$ but excluding trivial
cases, such $\rho_D$ necessarily presents negative eigenvalues
(Proposition \ref{pr:13}).

The quantum-mechanical overlap ($p$ and $1-p$) and localization ($\lambda$ and
$1-\lambda$) events are compatible and their combined probabilities can
computed, as shown in Table \ref{tab:1}. Similar tables exist for localization
probabilities involving more than two localization components. Similarly, the
trace inequalities can also be extended to the multi-component case.

The concavity property of the weight $\lambda$ implies some resemblance with the
same property of the quantum-mechanical (von Neumann) entropy. This is even more so for the quantity
$\cS=\log(\lambda)$, also a concave function of $\rho_A$, since under the
composition of systems $\lambda$ is (almost) multiplicative while $\cS$ is
additive, or more precisely super-additive.

Even if the weight $\lambda$ admits some interpretation as a probability, at
variance with the overlap probability $p$, it cannot be obtained as the
expectation value of an observable, at least not one independent of the very
state $\rho_A$ to be measured. This does not directly imply that $\lambda$ is
not a physical (or rather information-theoretic) quantity, since the same
holds for entropy, entanglement measure, and the like.

Finally, a possible cryptographic scheme is devised involving the localization
probability $\lambda$: The private state is masked by mixing it to produce a
public state; the private state can then be retrieved using the unique
decomposition along a subspace known only by the two private parties.

In Sec. \ref{sec:1}, precise mathematical definitions are provided,
the existence and uniqueness of the decomposition are proven (Theorem
\ref{th:1}) and many properties of it are uncovered and discussed in
detail for the finite-dimensional case, in particular its relation
with the Schur complement. The maximal property of the component $B$
is established (Lemma \ref{lm:1}), as is the concavity (Theorem
\ref{th:2}) and super-additivity attributes of the map $A\to B$ for
fixed $\cV$ (Proposition \ref{pr:11}), and the monotony of the map
$\cV \to B$ for fixed $A$ (Proposition \ref{pr:3}). Inequalities
related to the tensor product of spaces are also obtained (Proposition
\ref{pr:12}). Alternatively, the decomposition can be derived from the
orthogonal projector onto the subspace $A^{-1/2}\cV$ (Proposition
\ref{pr:9}) assuming here $A>0$ for simplicity, and also from the
orthogonality of the supports of $B$ and $C$ with respect to a
suitable $A$-dependent (but $\cV$-independent) new scalar product
(\eqs{2.25a} and \nec{2.25}). Using this idea, the decomposition
admits a natural extension to the case of $n>2$ localization
components. Still assuming $A>0$, the component $B$ is shown to be
given by ~$P A^{-1}P$ ~(Proposition \ref{pr:7}), being $P$ the
orthogonal projector operator onto $\cV$ (the general case $A\ge 0$ is
discussed in the main text). Several inequalities involving the trace
are also derived (Proposition \ref{pr:4} and Theorem \ref{th:3}), some
of them relying on the property $PAP \ge B$. Some of these results are
known to hold in infinite-dimensional Hilbert spaces as well.

%\newpage
\medskip\medskip
\appendix

\section{ Regularity of the map $\cV \to B(A|\cV)$ 
\label{app:a}}

All subspaces $\cV$ with dimension $n$ can be obtained as $\cV= U^{-1} \cV_0$
with a fixed $n$-dimensional subspace $\cV_0$ and a suitable unitary operator
$U$. Then, from clause 8 in the Proposition \ref{pr:10},
\be
\fTB(A|\cV)= \fTB(UAU^\dag|\cV_0)
.
\ee
If $U$ is expressed as $U = e^H$, with an anti-Hermitian $H$, the dependence
of $\fTB(A|\cV)$ on $\cV$ becomes a dependence on $H$. Such dependence is
real-analytic for $A>0$: $U$ is entire analytic as a function of (the matrix
elements of) $H$, and this translates to $UAU^\dag$. Moreover, the dependence
of $B=\fB(A|\cV)$ on (the matrix elements of) $A>0$ is meromorphic, with $B$
its Schur complement. The only singularities (with respect to any single
matrix element of $A$ as an independent variable) can be poles, however,
$0\le B \le A$ is bounded; therefore, there are no singularities as $H$
changes {\em within its anti-Hermitian} domain, so within that domain, the map
$H \to \fTB(A | e^H \cV)$ is real-analytic for $A>0$.

%\newpage

\acknowledgments The author thanks Carmen Garc{\'\i}a-Recio and
Enrique Ruiz Arriola for valuable discussions, and Miguel Mart{\'\i}n
Su\'arez for pointing out that Theorem \ref{th:1} is the Lebesgue
decomposition of positive operators, of which I was previously
unaware. This work has been partially supported by grant
PID2023.147072NB.I00 funded by MICIU/AEI/10.13039/501100011033 and by
ERDF/EU, and grant No. FQM-225 funded by Junta de Andaluc{\'\i}a.
\vfill

%\conflict interest
%\ethics statemtent
%\data access

\end{document}